\documentclass{article}

\usepackage{jheppub}

\usepackage[utf8]{inputenc}
\usepackage{verbatim}
\usepackage{natbib}
\usepackage{graphicx}
\usepackage{amsmath,amssymb}
\usepackage{bm}
\usepackage{dsfont}
\usepackage{cancel}
\usepackage{float}
\usepackage{braket}
\usepackage{siunitx}
\usepackage{xspace}
\usepackage[capitalize]{cleveref}
\usepackage{comment}
\usepackage{makecell}
\usepackage{xcolor}
\usepackage{caption}
\usepackage{subcaption}

\binoppenalty=\maxdimen
\relpenalty=\maxdimen 

\makeatletter
\gdef\@fpheader{\strut}
\makeatother

\crefname{section}{Sec.}{Secs.}

\newcommand*{\DM}{\ensuremath{\mathrm{DM}}}
\newcommand*{\SM}{\ensuremath{\mathrm{SM}}}
\newcommand*{\DR}{\ensuremath{\mathrm{DR}}}
\newcommand*{\nueff}{\ensuremath{\nu_{\mathrm{eff}}}}

\newcommand*{\mPl}{\ensuremath{m_\mathrm{Pl}}\xspace}

\newcommand*{\Neff}{\ensuremath{N_{\mathrm{eff}}\xspace}}

\newcommand*{\upd}{\mathrm{d}}

\newcommand{\ddN}[1]{\frac{\upd #1}{\upd N}}
\newcommand*{\rec}{\ensuremath{{\mathrm{rec}}}}

\title{
Superhorizon Isocurvature as a Window into Dark Matter Production
}

\author[a]{Christopher Gerlach,}
\affiliation[a]{%
 PRISMA$^{+}$ Cluster of Excellence \& Mainz Institute for Theoretical Physics,\\
 Johannes Gutenberg-Universit\"at Mainz, 55099 Mainz, Germany
}%
\emailAdd{cgerlach@uni-mainz.de}

\author[b]{Wolfram Ratzinger}
\affiliation[b]{Department of Particle Physics and Astrophysics,\\
Weizmann Institute of Science, Rehovot, Israel 7610001}
\emailAdd{wolfram.ratzinger@weizmann.ac.il}

\author[a]{and Pedro Schwaller}

\emailAdd{pedro.schwaller@uni-mainz.de}

\abstract{
In the presence of primordial isocurvature perturbations, for example in a separate dark radiation sector, the superhorizon evolution of curvature perturbations becomes nontrivial. If the dark sector is radiation-like and constitutes a significant fraction of the energy density, its isocurvature can imply isocurvature in the inflaton sector even without direct interactions between the sectors.
In this article, we revisit superhorizon curvature and isocurvature evolution in the long-wavelength limit systematically, drawing a simple picture of how to understand the nature of these fluctuations from first principles and without brute-force cosmic perturbation theory.
We show how the described setup is able to source isocurvature in simple models of dark matter such as freeze-in and freeze-out
and demonstrate that future measurements of matter and neutrino isocurvature can potentially discriminate between these two mechanisms.
}

\preprint{MITP-25-067}

\begin{document}

\maketitle

\newpage

\section{Introduction}

Cosmic perturbation theory has been remarkably successful in predicting the observables of the fiducial cosmological model, the $\Lambda\mathrm{CDM}$ model.
Providing a framework for the evolution of primordial fluctuations from their generation during inflation to the cosmic microwave background (CMB) anisotropies and the formation of large-scale structure (LSS), it lies in excellent agreement with high-precision data from, e.g., Planck \cite{Planck:2018vyg,Planck:2018jri} and galaxy surveys such as SDSS \cite{SDSS:2014iwm} and DESI \cite{DESI:2024mwx,DESI:2025zgx}. 

When solving the perturbed Einstein-Boltzmann equations in cosmic perturbation theory, initial conditions need to be chosen, specifically those of the perturbations. There exist two classes of initial conditions: \textit{Adiabatic} initial conditions are defined such that all components of the cosmic fluid, may it be photons, neutrinos, dark matter (DM), or even additional species like different types of dark radiation (DR), share the same density perturbations. In case of \textit{isocurvature} perturbations one or more of the constituents have deviating perturbations—commonly compared to the photons.

While observational data favors adiabatic initial conditions, percentage-level isocurvature contributions remain allowed.
Updated constraints stem from combinations of datasets such as CMB and the Lyman-$\alpha$ forest \cite{Chluba:2013dna,Planck:2018jri,Adshead:2020htj,Ramberg:2022irf,Buckley:2025zgh}. 
Even if subdominant, isocurvature remains of great interest: An observation of isocurvature would be a smoking gun for multi-field inflation. 
Moreover, small amounts of isocurvature may have an influence on the Hubble tension.
Last but not least, isocurvature perturbations could open a new window to study the dark sector: DM remains one of the greatest mysteries of physics and any hint for its properties will be helpful, may it be a single particle or a more complicated dark sector, including some form of dark radiation.

So far, DR and DM isocurvature have been studied extensively but usually in isolation.
DR isocurvature has been considered in the context of additional degrees of freedom \cite{Kawasaki:2011rc,DiValentino:2011sv,Kawakami:2012ke}, as a candidate for relaxing the Hubble tension \cite{Ghosh:2021axu,Chang:2025uvx}, and it has been considered in models of first-order phase transitions \cite{Freese:2023fcr,Elor:2023xbz,Buckley:2024nen}.
DM isocurvature on the other hand becomes particularly relevant for DM candidates which are sourced independently from the standard model sector, such as axions from the misalignment mechanism (see, for a review, \cite{Marsh:2015xka,OHare:2024nmr} and for recent updates on axion isocurvature, e.g., \cite{Berbig:2024ufe,Graham:2025iwx,Allali:2025pja}) or primordial black holes \cite{Papanikolaou:2020qtd,Domenech:2020ssp,Passaglia:2021jla,Kim:2025kgu,Gerlach:2025vco}. It has been found that freeze-in of particle DM, in contrast to thermal processes, would not fully dissolve potential initial isocurvature \cite{Bellomo:2022qbx} and the influence of adiabatic perturbations in the local rate during freezing-processes has been examined \cite{Holst:2023msh, Stebbins:2023wak}. Relations between matter and dark radiation isocurvature have been discussed in mechanisms linking the production of baryons, DM and additional radiation directly \cite{Berbig:2023uzs}.

In this work we focus on particle DM candidates which are produced from the standard model radiation bath via the freeze-in or freeze-out mechanism. Specifically, we will ask how the presence of DR isocurvature perturbations from an independent inflationary field may influence the DM curvature perturbations and what role the mechanism of DM production plays.

Our main result is that DM inevitably acquires isocurvature perturbations from DR, where the relative amplitude depends on the production mechanism and the energy contribution of the DR sector—even if the DR is only coupled gravitationally. To obtain this, we follow the separate universe approach \cite{Wands:2000dp} and derive simple evolution equations for superhorizon perturbations which reduce the problem to solving only the unperturbed background evolution numerically. We note that the imprints of freeze-in and freeze-out scenarios are differentiable.
Our findings demonstrate that isocurvature signatures would be able to help decipher the microphysics of DM. Therefore, future CMB and LSS measurements may already be able to distinguish between freeze-in and freeze-out scenarios based on the characteristic pattern they leave on DM isocurvature perturbations.

This paper is structured as follows: In \cref{sec:review} we review isocurvature, starting from the separate universe picture and the evolution of density perturbations. We then introduce definitions for the comoving curvature perturbation and isocurvature perturbations and derive evolution equations in the context of the separate universe approach. We also discuss the implications on adiabatic perturbations in the before introduced language.
In \cref{sec:emerging_isocurvature} we apply the results to different cosmological histories, including the decaying curvaton and DM production with freeze-in or freeze-out in presence of DR isocurvature. We then conclude in \cref{sec:results} how the imprint of the mechanism onto the isocurvature amplitudes could help to differentiate them. We give predictions for isocurvature observables in the basic cosmological scenarios.

\section{Review of isocurvature perturbations}
\label{sec:review}
\subsection{Separate universe approach}
Our goal is to investigate signatures of isocurvature perturbations on late time observables like the CMB fluctuations. This will require us to follow the evolution of fluctuations generated during inflation throughout the cosmic history, in order to use them as input when computing late time observables like the CMB anisotropies and LSS.

This task is highly simplified by the following observation: The fluctuations relevant for late time observables are far outside the horizon from the end of inflation to the emission of the CMB. Regions separated by the relevant length scales therefore can not influence each other, simply due to causality. We may then view the universe as a collection of patches, each of them evolving independently according to the same equations that govern the background dynamics. This method is known as the separate universe approach \cite{Wands:2000dp,Lyth:2003im,Lyth:2005fi,Artigas:2021zdk}.

Let us here take a first naive stab at this formalism. The universe is filled with an energy content. To keep our discussion concrete we will assume it consists of a number of fluids with energy densities $\rho_\alpha$ with the index labeling different fluids. Particle physics will specify how these fluids interact, and at the background level we will have to solve a differential equation of the following form
\begin{align}\label{eq: autonomous evolution}
	\dot{\bm{\rho}}=\bm{F}(\bm{\rho})~.
\end{align}
For example, for a single fluid with an equation of state $w$, the function $F$ would be given by $F(\rho)=-3(1+w)H\rho$, where $H^2=\rho/(3\mPl^2)$ denotes the Hubble rate. Note that this is an autonomous differential equation, since $\bm{F}$ can not explicitly depend on the time coordinate. One might therefore view all possible configurations $\bm{\rho}$ as a phase-space of the theory. In this case the solutions to the above differential equation can be regarded as trajectories, and for an initial value $\bm{\rho}_0$ at an initial time $t_0$ take the form
\begin{align}
	\bm{\rho}(t)=\bm{\rho}(\bm{\rho_0};t-t_0)~,\qquad \bm{\rho}(\bm{\rho_0};0)=\bm{\rho_0}~. \label{eq: solution to autonomous equation}
\end{align}
In the spirit of the separate universe approach, let us now consider two patches that are separated by a distance much larger than the horizon size. One patch starts with a perturbed initial condition $\bm{\rho}_0+\delta \bm{\rho}_0$. Consequently we may write the solution to the equation of motion to first order in $\delta \bm{\rho}_0$ as
\begin{align}
	\bm{\rho}(t)+\delta\bm{\rho}(t)\simeq\bm{\rho}(\bm{\rho_0}+\delta \bm{\rho}_0;t-t_0)~.\label{eq:perturbed background}
\end{align}
Inserting the left-hand side into \cref{eq: autonomous evolution} and separating the terms linear in $\delta\bm{\rho}$, one finds that the perturbation $\delta\bm{\rho}(t)$ is a solution to the initial value problem
\begin{align}
	\frac{d}{dt}\delta\rho_\alpha=\sum_{\beta}\delta\rho_\beta\frac{\partial F_\alpha}{\partial\rho_\beta}(\bm{\rho}(t))~,\qquad \delta\rho_\alpha(t_0)=\delta\rho_{0,\alpha}~. \label{eq: evolution delta rho} 
\end{align}
On the other hand we can find the perturbation by simply expanding the right-hand side of \cref{eq:perturbed background} and find
\begin{align}
	\delta\rho_{\alpha}(t)=\sum_{\beta}\frac{\partial \rho_{\alpha}(t)}{\partial\rho_{0,\beta}}\delta\rho_{0,\beta}~. \label{eq: delta rho from variation}
\end{align}
Voila! We have found the evolution equation of perturbations on super-horizon scales as well as the solution, which is simply obtained by varying the initial conditions of the background solution. So far we have however overlooked the coordinate dependence of $\delta\rho_{\alpha}$, an issue that we now need to fix.

\subsection{Curvature and isocurvature perturbations}
\label{sec: curvature and isocurvature perturbations}
\begin{figure}
    \centering
    \includegraphics[width=0.7\textwidth]{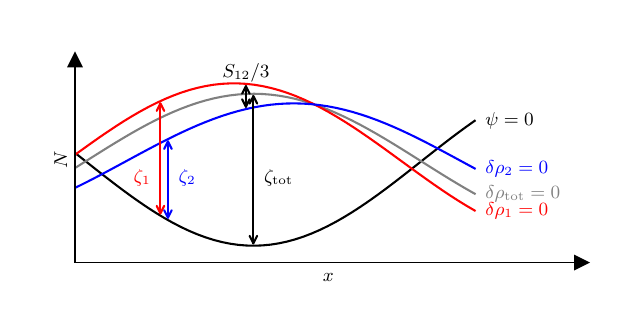}
    \caption{Geometric interpretation of the coordinate invariant perturbations as offsets in e-folds $\delta N$ (arrows) or equivalent time $\delta t=\delta N /H$ for spatially separated patches of the universe between slicings of constant density ($\delta \rho=0$) and the flat slicing with vanishing metric fluctuation ($\psi=0$).    
    }
    \label{fig:time_slices}
\end{figure}

Let us now proceed more carefully, and also include the fluctuations of the metric. This set of perturbations then features redundancies under coordinate transformations $t\rightarrow t+\delta t\,,\ x\rightarrow x+\delta x$\,. Since we are only interested in fluctuations on very large scales such that all quantities are in good approximation homogeneous, we must only concern ourselves with the former one. The density fluctuations transform under the time shift as
\begin{align}
    \delta \rho_{\alpha}\rightarrow \delta \rho_{\alpha}-\dot\rho_{\alpha} \delta t~. \label{eq:trafo_time_shift_density}
\end{align}
In order to find the relevant metric perturbations it is instructive to consider the transformation of the unperturbed metric
\begin{align}
    ds^2=-dt^2+a^2dx^2\rightarrow -\left(1-2\frac{\partial\delta t}{\partial t}\right)dt^2+a^2(1-2H\delta t)dx^2\,. \label{eq:trafo_time_shift_metric}
\end{align}
If we now consider a perturbed metric
\begin{align}
    ds^2=-(1-2\phi)dt^2+a^2(1-2\psi)dx^2\,,
    \label{eq:Newtonian_metric}
\end{align}
we see that $\psi$ transforms as  $\psi\rightarrow \psi+H\delta t$. 
This allows us to define \cite{Bardeen:1980kt,Bardeen:1983qw,Martin:1997zd}
\begin{align}
	\zeta_\alpha=-\psi-H\frac{\delta \rho_\alpha}{\dot\rho_\alpha}\, ,
\label{eq:definition_bardeen_potential}
\end{align}
which encodes the density perturbations in a coordinate invariant manner.
A rigorous definition of $\psi$ in terms of a general metric fluctuation and a derivation of all transformation properties may be found in \cref{app:full_metric}.

The constant time slices using the perturbed metric \cref{eq:Newtonian_metric} have an intrinsic 3-curvature proportional to the scalar perturbation $\psi$,
\begin{align}
    ^{(3)}R=\frac{4}{a^2}\nabla^2\psi~.
\end{align}
The curvature perturbation $\zeta_\alpha$ therefore gives the density fluctuation $\delta\rho_\alpha=-\dot\rho_\alpha \zeta_\alpha / H$ on a flat slicing with $\psi=0$ or, equivalently, the dimensionless curvature $\psi=-\zeta_\alpha$ on a constant density slicing $\delta\rho_\alpha=0$\,. 
Geometrically $\zeta_\alpha$ may therefore be regarded as the amount of time that separates these two slicings, as the transformation with $\delta t=\zeta_\alpha/H$ goes from the one with vanishing density fluctuation to the flat one.  Since $H=d\log(a)/dt$ specifies the evolution of e-folds $N=\log(a)$ with time, more directly $\zeta_\alpha$ gives the amount of e-folds $\delta N=\zeta_\alpha$ between these slicings, as illustrated in \cref{fig:time_slices}. 

Let us now return to the evolution of the fluctuation $\zeta_\alpha$. We have already found the solution for the perturbations $\delta \rho_\alpha$ in terms of background quantities in \cref{eq: delta rho from variation}. We now aim to do the same for $\psi$, making use of our intuition that it corresponds to the e-folds separating us from the flat slicing. This procedure is known as the $\delta N$-formalism\footnote{Later studies focused on higher orders of the expansion \cite{Tanaka:2006zp,Weinberg:2008nf,Weinberg:2008si,Takamizu:2010xy,Naruko:2012fe} and generalizing the concept \cite{Tanaka:2024mzw}.} \cite{Lyth:2004gb,Starobinsky:1982ee,Starobinsky:1985ibc,Sasaki:1995aw,Sasaki:1998ug,Tanaka:2010km} and may be justified more rigorously by the proof in \cite{Wands:2000dp}.
The amount of e-folds along a path may be found by solving the equation $\dot N=H$, leading to
\begin{align}
	N(t)=N(\bm{\rho_0};t-t_0)~,\qquad N(\bm{\rho_0};0)=0~. \label{eq: solution for Efolds}
\end{align}
In the following, we will jump from cosmic time to e-folds and back, depending on which picture is easier to understand. In general, they are completely interchangeable. The $\psi\simeq -\delta N$ along a perturbed path is then again given by variation with respect to the initial conditions $\bm{\rho_0}$,
\begin{align}
	\psi(t)=\psi_0-\sum_{\beta}\frac{\partial N(t)}{\partial \rho_{0,\beta}}\delta\rho_{0,\beta}\,.
\end{align}
Let us now put this result together with \cref{eq: delta rho from variation} in order to get the evolved $\zeta_\alpha$. Note that seemingly we have an extra initial condition in our equations, the $\delta\rho_{0,\alpha}$ and $\psi_0$, when only evolving the $\zeta_\alpha$. The additional initial condition corresponds to the arbitrary choice of the initial slicing. Without loss of generality we may choose it to be flat $\psi_0=0$ and $\delta\rho_{0,\alpha}=-\dot\rho_{0,\alpha}/H\, \zeta_{0,\alpha}$\,. We then find
\begin{align}\label{eq:evolution_zeta_general}
    \zeta_\alpha (t)&=\sum_{\beta} \left[ \frac{H(t)}{\dot\rho_{\alpha}(t)}\frac{\partial \rho_{\alpha}(t)}{\partial\rho_{0,\beta}} - 
    \frac{\partial N(t)}{\partial \rho_{0,\beta}}\right]
    \frac{\dot\rho_\beta(t_0)}{H(t_0)}\ \zeta_\beta(t_0)\,.
\end{align}
Although the different parts for the evolution of $\zeta_\alpha$ are well known, we could not find this closed expression in the literature. It has the particularly useful property that coordinate time $t$ might be exchanged for any other time $\tau$ such that $\tau(t;\mathbf{\rho}_0)$  is a coordinate transformation along the background trajectory as we show in \cref{app:time_coordinate_change}. This includes all combinations of $t,\rho_\alpha(t)$ and $N(t)$ that one may conceivably  use to parametrize the background evolution, e.g. the temperature of a relativistic species $T\sim \rho^{1/4}$\,. In this case, $H=\dot N$ is to be understood as a generalized Hubble rate with dots denoting derivatives with respect to $\tau$\,.  

The choice of e-folds $N$ as a time coordinate is particularly convenient, as the above equation simplifies to 
\begin{align}
    \zeta_\alpha (N)=\sum_{\beta} \frac{1}{\rho'_{\alpha}(N)}\frac{\partial \rho_{\alpha}(N)}{\partial\rho_{0,\beta}}\rho'_\beta(N_0)\ \zeta_\beta(N_0)~,
    \label{eq: zeta(N)}
\end{align}
where primes denote derivatives with respect to $N$\,.

Let us pause for a moment to recollect: We have found that the perturbations in energy density can be considered as a phase space. The evolution of density perturbations can be determined fully once the background, its evolution and the initial perturbations are known. This is inherited by the curvature perturbations. As can be seen from the equations of motion and the resulting equation for the curvature perturbation w.r.t. e-folds, i.e., \cref{eq: zeta(N)}, the curvature perturbation at $N$ can be fully determined by the initial curvature perturbations at $N_0$, if the evolution of the background densities is known.

This implies that it is possible to consider the evolution of $\zeta$ step by step, dividing the cosmic histories into sections with known background evolution. Since only the final perturbations matter for observations, the evolution of all curvature perturbations could be characterized by a product of transfer matrices, where each matrix compromises the evolution during one era. The transfer matrix approach has been used in \cite{Malik:2002jb} to evaluate the curvaton scenario, but in cosmic time.

So far, we have considered the curvature perturbation of each fluid by itself. Their linear combinations are however coordinate independent too and potentially more useful.
Let us therefore define the total curvature perturbation, which gives the e-folds that lie between a slicing of constant total energy density and the flat one,
\begin{align}\label{eq:zeta_total}
    \zeta_{tot}=\sum_\alpha \frac{\dot{\rho}_\alpha}{\dot{\rho}_{tot}}\zeta_\alpha~.
\end{align}
If two fluids have a relative difference in their respective curvature perturbation, we define this difference as the \textit{isocurvature} perturbation \cite{Wands:2000dp,Malik:2002jb}. For two fluids $\alpha \neq \beta$, it is given by
\begin{align}
    {S}_{\alpha \beta} \equiv 3(\zeta_{\alpha}-\zeta_{\beta})&= -3H \left( \frac{\delta\rho_{\alpha}}{\dot{\rho}_{\alpha}}- \frac{\delta\rho_{\beta}}{\dot{\rho}_{\beta}} \right),
    \label{eq:isocurvature_perturbation}\\
    &= -3 \left( \frac{\delta\rho_{\alpha}}{{\rho}_{\alpha}'}- \frac{\delta\rho_{\beta}}{{\rho}_{\beta}'} \right).
\end{align}

While there is a range of different formulations,\footnote{It can also be understood as relative entropy perturbation, since it can be rewritten in terms of entropy density $s$ and entropy density perturbations $\delta s$ in the respective fluids, $ S_{\alpha \beta}=\delta s_{\alpha}/s_{\alpha}-\delta s_{\beta}/s_{\beta}.$}
it often (when there is no further exchange of energy between the fluids) proves helpful to replace the time derivative with the continuity equation $\dot{\rho}_\alpha=-3H(1+w_\alpha)\rho_\alpha$ and then replace the perturbations by the density contrast $\delta_\alpha=\delta\rho_\alpha/\rho_\alpha$, which leads to
\begin{align}
    {S}_{\alpha \beta} = \frac{\delta_\alpha}{1+w_\alpha}-\frac{\delta_\beta}{1+w_\beta} \,.
    \label{eq:isocurvature_perturbation_delta}
\end{align}
Alternatively, instead of the $\zeta_\alpha$, $\zeta_{tot}$ together with a linearly independent subset of $S_{\alpha\beta}$ can be used to uniquely specify the perturbations.

\subsection{Implications for the evolution of adiabatic perturbations} 
The \textit{adiabatic} initial perturbation $\delta \rho_\alpha(N_0)=\rho'_\alpha(t_0) \delta N$, with a common $\delta N=H\delta t$ for all fluids, represents the special case where a patch is ahead by $\delta N$ with respect to the background, while following the exact same path through the phase space. This is equivalent to $\zeta_{tot}=-\delta N$ with all
\begin{align}
    {S}_{\alpha \beta} = -3 \left( \frac{\rho'_\alpha \delta N}{{\rho}_{\alpha}'}- \frac{{\rho}_{\beta}'\delta N}{{\rho}_{\beta}'} \right)=0\,.
\end{align}
The fact that the background evolution is autonomous implies that a patch which is ahead by $\delta N$ always stays ahead by this amount, as one can easily see from \cref{eq: solution to autonomous equation},
\begin{align}
    \bm{\rho}(\bm{\rho_0}+{\bm{\rho}}'(t_0)\delta N;N-N_0)=\bm{\rho}(\bm{\rho_0};N+\delta N-N_0)=\bm{\rho}(N)+{\bm{\rho}}'(N)\delta N\,.
\end{align}
In this way we recover the well known result that in the long-wavelength limit the adiabatic curvature perturbation does not source isocurvature and in the absence of isocurvature is conserved in magnitude. The common shift $\delta N$ or, respectively, $\delta t$ for all fluids is equivalent to the one-clock argument by Weinberg \cite{Weinberg:2004kr,Weinberg:2004kf}.

\section{Emerging isocurvature}
\label{sec:emerging_isocurvature}

\subsection{Cosmic histories}
\begin{figure}
    \centering
    \includegraphics[width=0.65\textwidth]{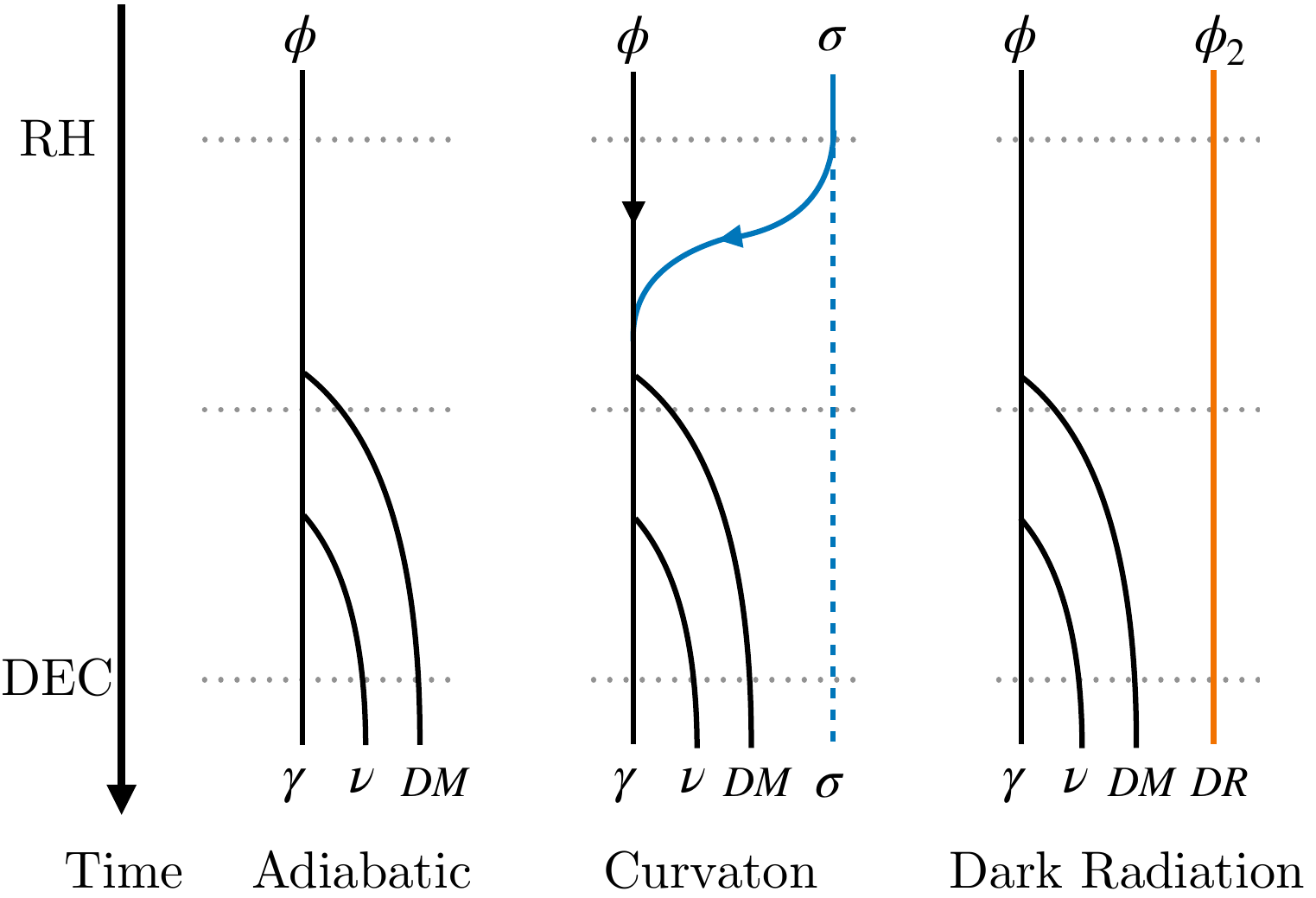}
    \caption{ 
    In the adiabatic scenario (\textbf{left}), the single inflaton $\phi$ reheats the universe. The time of inflaton reheating is marked by RH. All components of the cosmic fluid follow the same $\zeta$. It is inherited by the resulting DM, standard model neutrinos $\nu$ and photons $\gamma$, which are all decoupled at a later time (DEC).
    In the scenario of an additional decaying curvaton $\sigma$ (\textbf{center}), its energy density is completely transferred into the radiation sector, which is dominantly sourced by the inflaton field $\phi$. The decay of the curvaton starts after inflaton reheating, but in this case before DM decouples, resulting in purely adiabatic perturbations again. In the curvaton scenario, the inflaton carries zero $\zeta$.
    If we consider an additional inflationary field $\phi_2$ (\textbf{right}), which sources a dark sector consisting only of DR, the DM can inherit isocurvature by the presence of non-negligible DR curvature perturbations, even if DR couples only gravitationally to the other sector.}
\label{fig:cosmic_history_curvaton_decay}
\end{figure}

In Ref.~\cite{Gerlach:2025hxw} we investigated a number of cosmological histories and evaluated their potential for giving a non-zero neutrino isocurvature. Among the simplest models is a cosmology that features two fields during inflation. One of these reheats the standard model (SM) sector, while the second reheats a bath of free-streaming dark radiation. The term \textit{dark} refers here to the fact that this bath is only gravitationally coupled. DM and neutrinos decouple from the SM plasma at specific temperatures. We argued that since the neutrinos are in equilibrium with the SM at the time they decouple, they themselves feature no isocurvature. In terms of CMB observables the neutrinos are, however, nothing but free-streaming DR, which allows one to treat the sum of neutrinos and dark radiation as together as the \textit{effective} neutrinos. Since the DR stems from a different inflationary sector, these feature a non-zero isocurvature. This cosmic history is illustrated in \cref{fig:cosmic_history_curvaton_decay} (right).

If the sector featuring the DR also directly interacts with the DM, this would generically result in the DR inducing DM isocurvature. The induced DM isocurvature is fully correlated with the DR and, therefore, effective neutrino isocurvature.
The ratio between the two is hence constant over all patches and can be parametrized by the mixing angle $\varphi=\arctan(S_{\gamma\nu_{\rm eff}}/S_{\gamma\DM})$. We demonstrated that CMB and LSS observables depend on this angle and—in case isocurvature is discovered—allow this angle to be measured. 
Note that the correlation between the mixed isocurvature mode and the adiabatic mode in this scenario can be traced back to the correlation between the fluctuations in the two fields present during inflation. Such multi-field inflation models can in general realize arbitrary correlations between the two modes \cite{Gordon:2000hv}.

The main focus of this paper is to demonstrate that even in the minimal model, where the DR sector interacts purely gravitationally, an admixture of DM isocurvature is generated that depends on how the DM is generated from the SM bath, e.g. via freeze-in or freeze-out. We will do so from \cref{sec:analytic_system} onward by applying the previously established separate universe approach.

Before we want to illustrate the application of the separate universe approach on the well known example of the curvaton. This cosmology involves two fields during inflation as well. The inflaton field dominates the energy density at the end of inflation and reheats a radiation bath. The curvaton field starts oscillating at some point after inflation and redshifts like matter, such that its fraction of the total energy density can become sizable or even dominant. 
At some point the curvaton decays and further heats up the same radiation bath the inflaton already populated. In the end, there remains only energy in this radiation bath. All the fluids present at late times are generated from it. We illustrate this scenario in \cref{fig:cosmic_history_curvaton_decay} (center).

Since in this cosmology all energy is in one thermalized fluid after the decay of the curvaton, it can only feature adiabatic perturbations at late times. Initially, however, isocurvature between the inflaton and the curvaton may be present. Due to this isocurvature, the adiabatic perturbation is not conserved in this model at early times. Below we use the separate universe approach to calculate this induced curvature and show that we recover the same result as \cite{Malik:2002jb}, where the evolution of the perturbations was solved explicitly.

\subsection{The decaying Curvaton revisited}
\label{sec:decaying_curvaton}

Proposed more than two decades ago \cite{Mollerach:1989hu,Linde:1996gt,Moroi:2001ct,Enqvist:2001zp,Lyth:2001nq,Lyth:2002my}, the curvaton provides an alternative explanation for the generation of the observed adiabatic perturbations. In this scenario, the inflaton sector that initially dominates in energy and therefore the total (adiabatic) curvature has vanishing perturbations. Perturbations are, however, present in the subdominant curvaton that initially manifest themselves as isocurvature perturbations. 
As the energy in the curvaton becomes comparable to the inflaton sector, these then source an adiabatic perturbation. 
The scenario can be tested as it leads to enhanced non-Gaussianities, if the curvaton decays before dominating the total energy \cite{Lyth:2002my,Planck:2019kim}. Further the scenario leads to large remnant isocurvature if the curvaton decays after or simultaneously with the decoupling of other fluids, e.g. DM \cite{Lyth:2002my,Lyth:2003ip,Gupta:2003jc}. In the simple case that we study here, where after the curvaton decay only one thermalized fluid is present, this effect is, however, absent.

\begin{figure}
\centering
\begin{subfigure}{.5\textwidth}
  \centering
  \includegraphics[width=.9\linewidth]{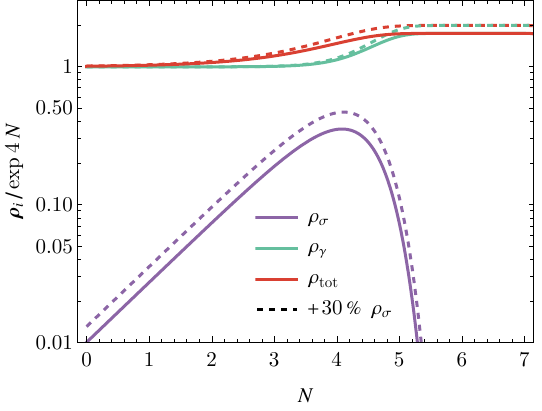}
  \caption{$\Tilde{\Gamma}_{0}=10^{-3}$.}
\end{subfigure}%
\begin{subfigure}{.5\textwidth}
  \centering
  \includegraphics[width=.9\linewidth]{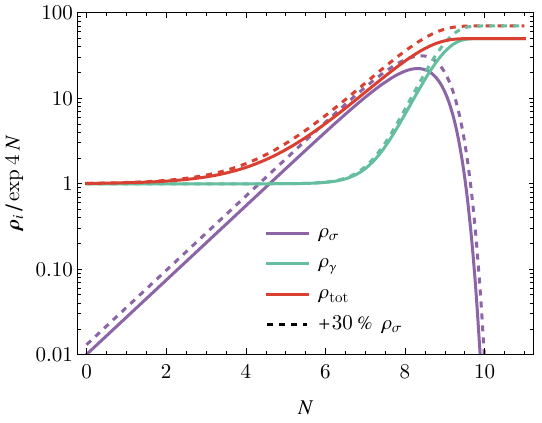}
  \caption{$\Tilde{\Gamma}_{0}=10^{-6}$.}
\end{subfigure}
\caption{Background evolution of the direct curvaton decay for two different decay rates. We show the energy density for the curvaton (purple), the radiation sector (green) and the total energy density (red). The dashed line is the same setup with a 30 percent higher initial curvaton abundance, resembling a patch with a density perturbation in the curvaton. Due to the matter scaling of the curvaton, the total energy density normalized to the radiation energy density increases. The smaller decay constant in (b) causes the curvaton to dominate the energy density before decaying to radiation.}
\label{fig:numerical_evolution_curvaton_decay_background}
\end{figure}

Let us now solve for the perturbations using the separate universe approach. As stated before, the curvaton field $\sigma$ starts with subdominant energy density, but carries the only non-vanishing perturbation $\zeta_{\sigma,0}$. The SM radiation sector $\gamma$, previously reheated by the inflaton $\phi$, dominates the energy density, but inherited zero curvature perturbation from the inflaton. The initial perturbations therefore are
\begin{align}
    \zeta_{\gamma}(N_0)=0 \,,&\quad \zeta_{\sigma}(N_0)=\zeta_{\sigma,0} \,,\\
    \zeta_{\text{tot}}(N_0)\approx\zeta_{\gamma}(N_0)=0\,,&\quad \mathcal{S}_{ \gamma\sigma}(N_0)=-3\zeta_{\sigma,0}\,.
\end{align}

Once the oscillation scale of the curvaton enters the horizon, it behaves like matter. We assume that at the same time it decays with a fixed rate $\Gamma$ with respect to cosmic time into light particles of the radiation sector $\gamma$, therefore further heating it up.
The equations of motion in terms of e-folds for the background fields become
\begin{align}
    \rho'_\sigma&=-(3+\Tilde{\Gamma})\rho_\sigma\,, \\
    \rho'_\gamma&=-4\rho_\gamma+\Tilde{\Gamma}\rho_\sigma \,,
 \end{align}
where the prime denotes the derivative with respect to e-folds $N$ and $\Tilde{\Gamma}=\Gamma/H$ is the interaction rate adjusted to the e-fold parameterization. Plugging the initial conditions, \cref{eq: zeta(N)} then reads for the two fluids
\begin{align}\label{eq:curvaton curvature}
    \zeta_\sigma(N)=\frac{1}{\rho'_\sigma(N)}
    \frac{\partial \rho_\sigma}{\partial\rho_{\sigma,0}}\rho'_\sigma(N_0)\zeta_{\sigma,0}\,,  \quad
    \zeta_\gamma(N)=\frac{1}{\rho'_\gamma(N)}
    \frac{\partial \rho_\gamma}{\partial\rho_{\sigma,0}}\rho'_\sigma(N_0)\zeta_{\sigma,0}~.
\end{align}
The above equations imply that we can find the evolution of the perturbations by considering how the solution of the background depends on its initial conditions. In the simplified case of $\Gamma=0$, we may even solve the background equations analytically to find $\rho_\sigma=\rho_{\sigma,0}\exp(-3N)$ and $\rho_\gamma=\rho_{\gamma,0}\exp(-4N)$\,. We can then trivially carry out the variations with respect to the initial conditions in the equations above and find $\zeta_\sigma(N)=\zeta_{\sigma,0}$ and $\zeta_\gamma(N)=0$\,. \cref{eq:zeta_total} then, however, implies that the total curvature perturbation $\zeta_{\text{tot}}$ becomes non-zero as the universe goes from radiation to curvaton dominated and eventually asymptotes to $\zeta_{\sigma,0}$
\begin{equation}
    \zeta_{\text{tot}}(N)=\frac{3\rho_\sigma(N)}{3\rho_\sigma(N)+4\rho_\gamma(N)}\zeta_{\sigma,0}\,.
\end{equation}
Let us again stress that the calculation turned out this simple due to two factors: First, the use of the separate universe approach saved us the effort of finding evolution equations for the perturbations and solving them. Second, since we worked with e-folds $N$ as our time coordinate, we could trivially ensure that while varying $\rho_{\sigma,0}$ along all paths the same amount of e-folds passed. When working with a different time coordinate, one has to take the variation of e-folds with the initial conditions into account according to \cref{eq:evolution_zeta_general} as was done in \cite{Wands:2000dp} to arrive at the same result. For more complicated models, usage of a special time coordinate might significantly help when solving the background equations, in which case including the variation of e-folds might be the easiest option.

\begin{figure}
\centering
\begin{subfigure}{.5\textwidth}
  \centering
  \includegraphics[width=.9\linewidth]{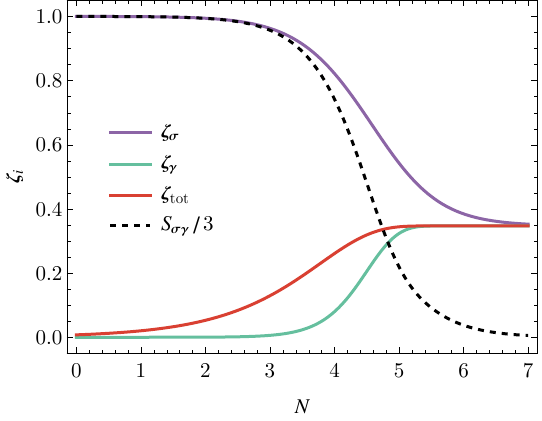}
  \caption{$\Tilde{\Gamma}_{0}=10^{-3}$.}
  \label{fig:sub1}
\end{subfigure}%
\begin{subfigure}{.5\textwidth}
  \centering
  \includegraphics[width=.9\linewidth]{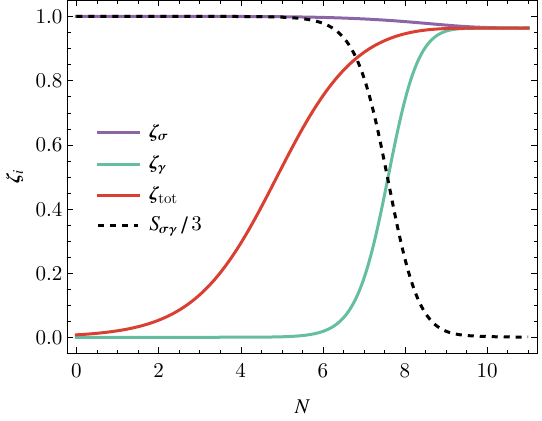}
  \caption{$\Tilde{\Gamma}_{0}=10^{-6}$.}
  \label{fig:sub2}
\end{subfigure}
\caption{The numerical evolution of curvature and isocurvature perturbations is shown. The initial curvature of the curvaton field (purple) is normalized to unity. The radiation (green) starts with zero curvature. The isocurvature perturbation (black) is divided by 3. The total curvature perturbation is depicted as well (red). As can be seen from the background evolution in \cref{fig:numerical_evolution_curvaton_decay_background}, in the case of a small decay constant the matter-scaling curvaton sector can dominate the energy density before its decay, thereby leaving a stronger imprint of its associated curvature perturbations.}
\label{fig:numerical_evolution_curvaton_decay}
\end{figure}

Let us now turn back to the more complicated case of the decaying curvaton, where we need to solve the background equations numerically. We consider the two benchmarks from Ref.~\cite{Malik:2002jb}, $\Tilde{\Gamma}_0=10^{-3}$ and $\Tilde{\Gamma}_0=10^{-6}$, respectively. The initial total energy density is normalized to one, specifically we start with $\rho_{\gamma,0}=0.99$ and $\rho_{\sigma,0}=0.01$. We show the background evolution of both energy densities in \cref{fig:numerical_evolution_curvaton_decay_background} as solid lines. 

To find the evolution of the perturbations, we again need to consider patches that start with a perturbation in $\rho_{\sigma,0}$\,. For illustration purposes we show the evolution of such patches that start with a  30\% increased $\rho_{\sigma,0}$ as dashed lines in \cref{fig:numerical_evolution_curvaton_decay_background}. We find that in the first scenario, where the curvaton decays before dominating, this leads to about a 10\% increase in the final total energy, while it leads to almost a full 30\% increase if the curvaton dominates before decaying as is the case in the second example. 

To find the evolution of the  perturbations, we now numerically take the increase in $\rho_{\sigma,0}$ to be very small. This allows us to evaluate the derivatives in \cref{eq:curvaton curvature} and gives us the curvature for every $N$. We show the evolving perturbations in \cref{fig:numerical_evolution_curvaton_decay}. The fact that an increase of the curvaton density leads to almost the same increase in the final radiation and total density in the second scenario is now captured by the perturbations tending to 1, while in the first scenario, where the effect is suppressed, they go to $\approx1/3$\,.
For the lower decay rate, it takes longer for the decay process to become efficient. The curvaton has more time to contribute stronger to the energy density and the imprint of its curvature is enhanced. Notably, we successfully reproduce the result from \cite{Malik:2002jb} without deriving evolution equations for the perturbations and solving them.
Beware that since we are only working at linear order in the perturbations, we can in principle choose any initial value for the fluctuations to find their transition functions. They can later be scaled to match the primordial values from observations.

The decaying curvaton scenario provides us with an important insight: Despite starting with isocurvature between curvaton and radiation, the universe in this scenario ends up with zero isocurvature, once the curvaton completely decayed into radiation—thereby sourcing the observed adiabatic curvature perturbations. The universe can be born non-adiabatic, but still die adiabatic.

\subsection{Analytic toy model for DM production}
\label{sec:analytic_system}

Let us now take a closer look at a decoupling process between two fluids $\rho_1$ and $\rho_2$ with vanishing initial isocurvature $S_{12}=0$ in the presence of a third fluid $\rho_3$ which is only gravitationally coupled. This situation corresponds e.g. to the upper half of the right example in \cref{fig:cosmic_history_curvaton_decay}, where $\rho_1$ and $\rho_2$ may correspond to the SM plasma and DM, while $\rho_3$ is DR. We are interested in the final isocurvature between the first two fluids in the case that there is an initial isocurvature between them and the third fluid, $S_{13}\neq 0$.

For now we study a simplified system of fluids that mimics dynamics typically found in DM freeze-in and freeze-out processes but does not exactly correspond to any particle physics theory. We assume that all three fluids red-shift like radiation and that between fluids one and two there is a heat exchange $Q$\,. In terms of e-folds $N$ the background evolution is given by
\begin{align}
    \rho_1'&=-4\rho_1-Q/H  \\
    \rho_2'&=-4\rho_2+Q/H  \\
    \rho_3'&=-4\rho_3\,,
    \label{eq:EoM_toy_emerging_isocurvature}
\end{align}
where $H=\sqrt{(\rho_1+\rho_2+\rho_3)/(3\mPl^2)}$ is the Hubble rate and primes denote derivatives with respect to $N$. The rate $Q$ is supposed to model a particle physics interaction and it should therefore only depend on the fluids that it involves $Q=Q(\rho_1,\rho_2)$\,. To keep the solution analytic, we assume that it takes the following form
\begin{align}
    Q=\left( f_{\mathrm{eq},2}\rho_1-f_{\mathrm{eq},1}\rho_2\right)\Gamma_\mathrm{crit}\frac{\rho_1+\rho_2}{\rho_\mathrm{crit}+\rho_1+\rho_2}\,.
\end{align}
The interaction drives fluids one and two to an equilibrium, where they respectively constitute fractions $f_{\mathrm{eq},1}$ and $f_{\mathrm{eq},2}$ of the total energy in the sector $\rho_1+\rho_2$\,. Therefore, $f_{\mathrm{eq},1}+f_{\mathrm{eq},2}=1$.
The interaction is most efficient when $\rho_\mathrm{crit}\sim\rho_1+\rho_2$ and takes place at a rate $\Gamma_\mathrm{crit}$\,. In terms of particle physics one may think of the equilibrium state being determined e.g. by the amount of relativistic degrees of freedom in fluids one and two, with the interaction rate coming from some portal that eventually gets quenched. 

The solutions for $\rho_1+\rho_2$ and $\rho_3$ are given by 
\begin{align}
    \rho_1+\rho_2=\left(\rho_{1,0}+\rho_{2,0}\right)\exp(-4N)\,,\quad \rho_3=\rho_{3,0}\exp(-4N)\,.
\end{align}
It is convenient to define $N_\mathrm{crit}$ via $(\rho_{1,0}+\rho_{2,0})e^{-4N_\mathrm{crit}}=\rho_\mathrm{crit}$. This allows us to rewrite the evolution of $\rho_1$ as
\begin{align}
    \rho_1'&=-4\rho_1-\left(\rho_1-f_{\mathrm{eq},1}\rho_\mathrm{crit}e^{-4(N-N_\mathrm{crit})}\right)\frac{\Gamma_\mathrm{crit}}{2H_\mathrm{crit}}\frac{1}{\cosh(2(N-N_\mathrm{crit}))}\,,
\end{align}
where $H_\mathrm{crit}=H_0 e^{-2N_\mathrm{crit}}$ is the Hubble rate at $N_\mathrm{crit}$ and $H_0=\sqrt{(\rho_{1,0}+\rho_{2,0}+\rho_{3,0})/(3\mPl^2)}$ the initial one.  If the initial conditions are chosen such that $\rho_{1,0}+\rho_{2,0}\gg\rho_\mathrm{crit}$ and therefore $N_\mathrm{crit}\gg 1$, we may approximate the solution as
\begin{align}
    \rho_1=\rho_\mathrm{crit}e^{-4(N-N_\mathrm{crit})} \left(f_{\mathrm{eq},1} + (f_{0,1}-f_{\mathrm{eq},1})e^{-\frac{\Gamma_\mathrm{crit}}{2H_\mathrm{crit}}\arctan\left(e^{2(N-N_\mathrm{crit})}\right)}\right)\,,
\end{align}
where $f_{0,1}=\rho_{1,0}/(\rho_{1,0}+\rho_{2,0})$\,. The solution for the second fluid may be cast into an analogous form. With this analytic solution in hand, we may again determine the evolution of the perturbations with the help of \cref{eq: zeta(N)} by variation with respect to the initial values $\rho_{i,0}$, upon which $N_\mathrm{crit}\,,\ H_\mathrm{crit}$ and $f_{0,i}$\, implicitly depend.

\textbf{Freeze-In.}
Let us now take a closer look at a more specific scenario that corresponds to the second fluid freezing in. For this purpose we assume $\rho_{2,0}=0$ and $f_{\mathrm{eq},2}\neq0$ such that around $N_\mathrm{crit}$ the interaction populates the second fluid. The resulting isocurvature $S^\mathrm{fin}_{12}$ once the interaction seizes to be efficient at $N\gg N_\mathrm{crit}$ is given by
\begin{align}\label{eq:analytic_FI}
    S^\mathrm{fin}_{12}&=-\frac{\pi}{8}\frac{\Gamma_\mathrm{crit}}{H_\mathrm{crit}}\frac{1}{1-e^{-\frac{\pi}{4}\frac{\Gamma_\mathrm{crit}}{H_\mathrm{crit}}}}\frac{1}{1+f_{\mathrm{eq},1}\left(e^{\frac{\pi}{4}\frac{\Gamma_\mathrm{crit}}{H_\mathrm{crit}}}-1\right)}\Omega_3 S^{0}_{13}\\
    &\approx-\Omega_3 S^{0}_{13}
    {\begin{cases}
			\frac{1}{2}, & \Gamma_\mathrm{crit}\ll H_\mathrm{crit}\\
            \frac{\pi}{8}\frac{\Gamma_\mathrm{crit}}{H_\mathrm{crit}}, &H_\mathrm{crit}\ll\Gamma_\mathrm{crit}\ll -\frac{4}{\pi}\log\left(f_{\rm eq,1}\right)H_\mathrm{crit}\\
            \frac{\pi}{8}\frac{\Gamma_\mathrm{crit}}{H_\mathrm{crit}}\frac{e^{-\frac{\pi}{4}\frac{\Gamma_\mathrm{crit}}{H_\mathrm{crit}}}}{f_{\mathrm{eq},1}}, & -\frac{4}{\pi}\log\left(f_{\rm eq,1}\right)H_\mathrm{crit}\ll\Gamma_\mathrm{crit}
		 \end{cases}}\,,
\end{align}
in terms of the initial isocurvature $S^{0}_{13}$\,. Remarkably, isocurvature is induced in the sector of fluids 1 and 2 by the third fluid that is, however, only gravitationally coupled. In \cref{subsec:FI_numerical} we will study a similar system, where the first fluid is the SM radiation bath, the third is DR and the second fluid which freezes in is DM.

We may interpret the result in the two limits of a weak and a strong interaction in the following way. Note that both at times before and after the interaction is efficient, the isocurvature $S_{ij}$ corresponds to perturbations in $\log({\rho_i}/{\rho_j})$\,,
\begin{align}
    \delta\log\left(\frac{\rho_i}{\rho_j}\right)=\left(\frac{\delta\rho_i}{\rho_i}-\frac{\delta\rho_j}{\rho_j}\right)=\frac{4}{3}S_{ij}\,.
\end{align}
In terms of the separate universe approach, a positive $S^{0}_{13}$ therefore corresponds to a patch in which the density $\rho_3$ is reduced with respect to $\rho_1+\rho_2$\,. In this patch the Hubble rate is always smaller at a given $\rho_1+\rho_2$ and therefore interaction rate $Q$. This results in that patch spending more time at a given interaction rate and therefore an enhanced yield $\rho_2/\rho_1$ at late times compared to the background patch, explaining the sign of the isocurvature.

Since the third fluid is only gravitationally coupled, the dependence of the yield on $\rho_3$ is only due to the Hubble rate, which is why the result is proportional to $\Omega_3$\,. While in the case of a weak interaction the time spent at a given rate is the main factor determining the yield, for strong interactions the yield eventually tends towards the equilibrium $\rho_2/\rho_1=f_{\mathrm{eq},2}/f_{\mathrm{eq},1}$ and becomes independent of the Hubble rate and therefore $\rho_3$\,. This explains the exponential suppression of the isocurvature $S^\mathrm{fin}_{12}$ at large interaction strength. Of course the region of parameter space where the DM thermalizes with the SM is anyways not relevant for freeze-in DM.

\textbf{Freeze-Out.}
Within in the same toy model we can also study the case in which the first fluid freezes out. To mimic this scenario, assume that prior to the dynamics fluids one and two were in thermal equilibrium such that the ratio $\rho_{1,0}/\rho_{2,0}$ is set to some fixed value. This implies that $S^{0}_{12}=0$\,. The dynamics described above then tend to diminish the first fluid, if we choose $f_{\mathrm{eq},1}=0$\,. In this case the final isocurvature in terms of $S^{0}_{13}=S^{0}_{23}$ is given by
\begin{align}
    S^\mathrm{fin}_{12}&=-\frac{\pi}{8}\frac{\Gamma_\mathrm{crit}}{H_\mathrm{crit}}\frac{1}{1-f_{0,1}e^{-\frac{\pi}{4}\frac{\Gamma_\mathrm{crit}}{H_\mathrm{crit}}}}\Omega_3 S^{0}_{13}\\
    &\approx-\frac{\pi}{8}\frac{\Gamma_\mathrm{crit}}{H_\mathrm{crit}}\Omega_3 S^{0}_{13}
    \begin{cases}
			\frac{1}{1-f_{0,1}}\,, & \Gamma_\mathrm{crit}\ll H_\mathrm{crit}\\
            1\,, & \Gamma_\mathrm{crit}\gg H_\mathrm{crit}
		 \end{cases}\,,
\end{align}
Conversely to the previous example the ratio $\rho_{1}/\rho_{2}$ is now by construction fixed at the initial value if the interaction is weak, which leads to the parametric dependence on ${\Gamma_\mathrm{crit}}/{H_\mathrm{crit}}$\,. In the limit where ${\Gamma_\mathrm{crit}}/{H_\mathrm{crit}}$ is large, the system tends to the new equilibrium $\rho_1=0$. The tiny rest that remains is all that is needed to explain the observed DM abundance in freeze-out models. This results in an unsuppressed isocurvature in this case. As we will see in \cref{subsec:FO_numerical}, this resembles a system where the fluid $\rho_2$ will be the SM radiation bath, the third will be DR and the $\rho_1$ here will be the freeze-out DM.

\subsection{Numerical analysis of freeze-in}
\label{subsec:FI_numerical}

The dark matter freeze-in process \cite{Hall:2009bx,Chu:2011be,Chu:2013jja,Blennow:2013jba,Bernal:2017kxu,DEramo:2017ecx} is a mechanism to create DM with much smaller required couplings than the freeze-out scenario. As a consequence, the DM particle is never in thermal equilibrium with the standard model and is rather produced via rare processes from bath particles. With decreasing temperature, the production channels become Boltzmann suppressed and the abundance \textit{freezes in}. Freeze-in can be differentiated into UV- and IR-freeze-in \cite{Cirelli:2024ssz}. We focus on the later case here.

In the analytic toy model we have seen that the process of DM freeze-in can source isocurvature between the SM and the DM fluids in the presence of a third fluid with non-vanishing isocurvature. Let us now investigate a realistic interaction term with numerical methods.
As before, the system contains three fluids: The SM photons with energy density $\rho_\gamma$, DM with $\rho_\DM$ and a generic dark radiation sector $\rho_\DR$, where the DM abundance is negligible initially.

The SM bath is in thermal equilibrium with a heavy particle $B_1$ of mass $m_B$ and with $g_B$ degrees of freedom. This particle decays into a particle $B_2$, which is in equilibrium with the SM, and a DM particle X, $B_1\rightarrow B_2 + X$, with a constant rate $\Gamma$. The corresponding interaction term in the Boltzmann equation is computed in \cite{Hall:2009bx}. Adapted to the language introduced here, this interaction reads
\begin{align}
    Q_{\gamma \to\DM}=\frac{g_{B}}{2\pi^2}\,\Gamma m_B^2 m_X T \,K_1\left(m_B/T \right),
    \label{eq:Q_freeze-in}
\end{align}
with $K_1$ being the Bessel function of second type, $m_B$ the mass of $B_1$ and $m_X$ the DM mass.
The temperature is that of the SM bath, $T=(30\rho_{\gamma}/(\pi^2g_\star))^{1/4}$. The interaction enters the equations of motion as above in \cref{eq:EoM_toy_emerging_isocurvature}, where the second component resembling DM now redshifts accordingly,
\begin{align}
    \rho_\gamma'&=-4\rho_\gamma-Q/H\,,  \\
    \rho_\mathrm{DM}'&=-3\rho_\mathrm{DM}+Q/H\,,  \\
    \rho_\mathrm{DR}'&=-4\rho_\mathrm{DR}\,.
    \label{eq:EoM_dark_matter}
\end{align}
The energy density is transferred from the SM radiation bath towards the freeze-in DM. The photon bath as well as DR redshift as radiation.

We solve the system numerically and study the evolution of the perturbations. To achieve a stable solution, the numerical evolution starts with negligible but finite DM abundance. Initially we set the isocurvature between radiation bath and DM to zero, $\zeta_{\gamma,0}=\zeta_{\DM,0}=0.1$ such that $S^0_{\gamma\DM}=0$. The DR sector starts with $\zeta_{\DR,0}=-0.1$, so initially the isocurvature is given by $S_{\gamma\DR,0}=3(\zeta_{\gamma,0}-\zeta_{\DR,0})=0.6$\,.

\begin{figure}[t]
\centering
\begin{subfigure}{.5\textwidth}
  \centering
  \includegraphics[width=.9\linewidth]{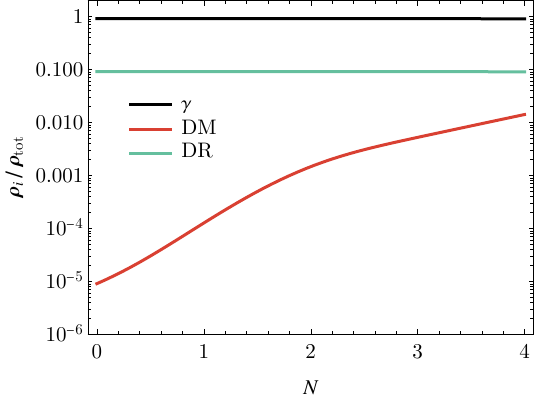}
  \caption{Freeze-in background evolution.}
  \label{fig:curv_evo_FI:sub1}
\end{subfigure}%
\begin{subfigure}{.5\textwidth}
  \centering
  \includegraphics[width=.9\linewidth]{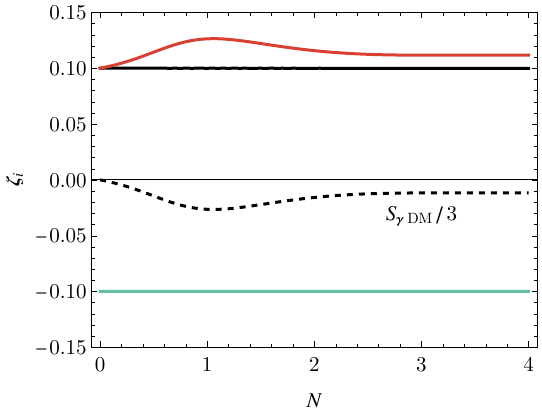}
  \caption{Freeze-in curvature perturbations.}
  \label{fig:curv_evo_FI:sub2}
\end{subfigure}
\hfill
\vspace{2pt}
\begin{subfigure}{.5\textwidth}
  \centering
  \includegraphics[width=.9\linewidth]{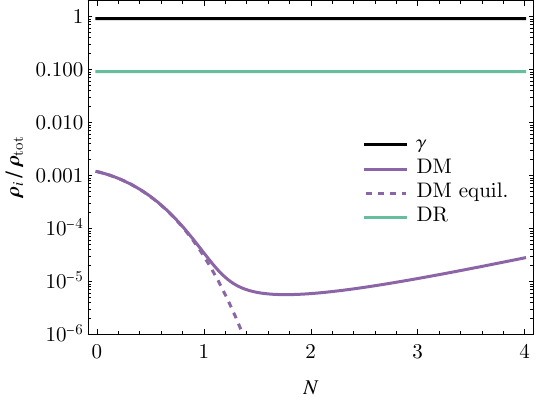}
  \caption{Freeze-out background evolution.}
  \label{fig:curv_evo_FO:sub1}
\end{subfigure}%
\begin{subfigure}{.5\textwidth}
  \centering
  \includegraphics[width=.9\linewidth]{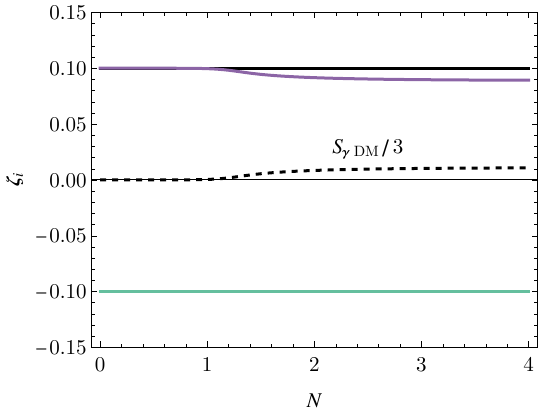}
  \caption{Freeze-out curvature perturbations.}
  \label{fig:curv_evo_FO:sub2}
\end{subfigure}
\caption{Evolution of background energy densities and curvature perturbations in terms of e-folds for freeze-in (top, DM in red) and freeze-out (bottom, DM in purple). The color is corresponding to the fluids. In the right panels, the black dashed line is the isocurvature perturbation between the SM and the DM sector. The initial DR energy density is the same for the two presented setups. In the freeze-out, the equilibrium energy density is included as dashed purple line. 
}
\label{fig:curv_evo}
\end{figure}

The resulting evolution of the background densities and the curvature perturbations is shown in \cref{fig:curv_evo_FI:sub1} and \cref{fig:curv_evo_FI:sub2}, respectively.  We show the SM sector (black), dark radiation (green) which starts with the contribution $\rho_{\DR,0}=0.1\,\rho_{\gamma,0}$, and the freeze-in dark matter (red). The dashed line in the right panel is the isocurvature between SM and DM computed from both curvature perturbations. More details on the chosen numerical parameters and units can be found in \cref{sec:app:numeric_evol}.

In the left panel, we can see the DM density increasing faster than pure matter scaling while it is produced from the SM bath. Between $N=1$ and 2, when $T\sim m_B$, the production becomes inefficient and the DM abundance is \textit{frozen in}. It starts scaling as matter.
We find that during the freeze-in process, $\zeta_{\mathrm{DM}}$ is increased, leading to a finite negative isocurvature perturbation as our intuition from the analytic toy model would suggest (see \cref{fig:curv_evo_FI:sub2}). The perturbations \textit{freeze} together with the DM abundance at around $N\approx 2$, when the DM production ends. In total, we are left with a finite isocurvature perturbations between the $\gamma$ bath and DM, ${S}_{\gamma\DM}$. 

\begin{figure}[t]
\centering
\begin{subfigure}{.5\textwidth}
  \centering
  \includegraphics[width=.9\linewidth]{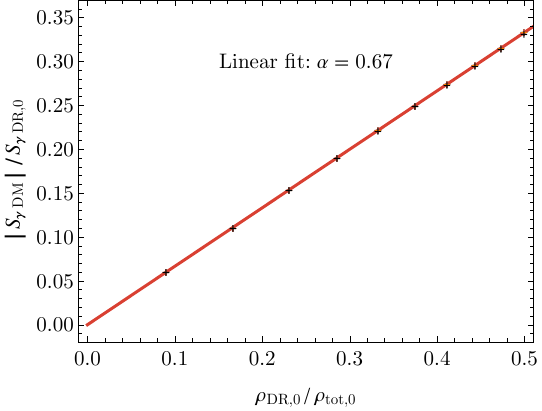}
  \caption{Freeze-in.}
  \label{fig:scan_sub1}
\end{subfigure}%
\begin{subfigure}{.5\textwidth}
  \centering
  \includegraphics[width=.9\linewidth]{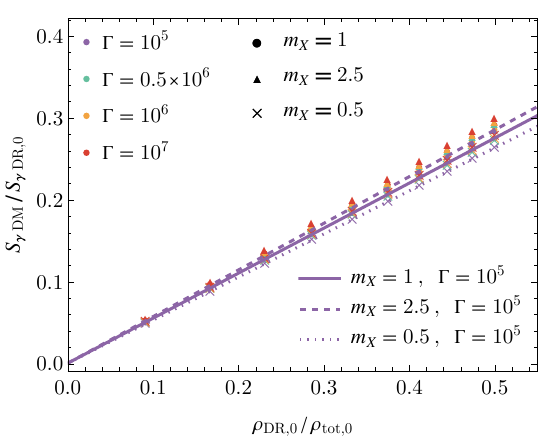}
  \caption{Freeze-out.}
  \label{fig:scan_sub2}
\end{subfigure}
\caption{We show the dependence of the final isocurvature of dark matter, normalized by the initial isocurvature of dark radiation, as a function of the dark radiation energy contribution. The left panel shows the freeze-in, where the results for different rates coincide. The red line is a linear fit. In the right panel, the same is done for freeze-out scenarios with different interaction rates and DM masses $m_X$. The purple lines are fits to the points corresponding to one mass each with $\Gamma=10^5$. While the general linear dependence on the DR contribution remains, the toy model parameters still have a small influence on the resulting isocurvature.}
\label{fig:scan_fit}
\end{figure}

An analysis of the parametric dependence is shown in the left panel of \cref{fig:scan_fit}. For this, we perform the numerical evolution multiple times (see also \cref{fig:yield_combi}), varying the model parameters as well as the initial DR contribution, and read out the final values at the end of the evolution, once they have converged. We plot the resulting isocurvature perturbation in DM at the end of the evolution, normalized by the initial isocurvature of DR, against the initial DR energy contribution. 
Here it becomes obvious that the scaling is strictly linear, as the interaction of the DR sector is purely gravitational,
\begin{align}
    S_{\gamma\DM}=\alpha\,\Omega_{\DR,0}\,S_{\gamma\DR,0}\,.
    \label{eq:slope_FI}
\end{align}
where a linear fit reveals the slope $\alpha=-0.67$.

This result agrees very well with our intuition: The more the DR sector contributes to the energy density during the freeze-in, and the larger its isocurvature amplitude is, the more significant the imprint into the DM isocurvature perturbation becomes. If there is zero initial isocurvature in the DR, we cannot source DM isocurvature. The sign results from the fact that in freeze-in, the DM abundance increases if one lowers the amount of DR and with it the Hubble rate.

Different decay rates $\Gamma$ and the time of freeze-in, determined by the  mass $m_B$, have no influence on the resulting isocurvature. 
The dependence therefore is as predicted in the analytic toy model (cf. \cref{eq:analytic_FI} for small rates). Only the prefactor in the analytic model deviates slightly. In \cref{app:sudden_freeze} we calculate the factor in the sudden freeze-in approximation introduced in \cite{Lyth:2003ip} and find $-2/3$ in excellent agreement with the numerical study. There it seems that the difference is mainly due to DM redshifting as $\propto a^{-3}$, while $\rho_2$ in the toy model was redshifting like radiation $\propto a^{-4}$\,. Note that in reality the DM particles produced in the decay of $B_1$ have a residual momentum and would show a behavior in between these two cases. Investigation of this effect as well as UV freeze-in, where DM production has to be discussed together with the reheating of the SM sector, are left to future work.

\subsection{Numerical analysis of freeze-out}
\label{subsec:FO_numerical}

The numerical analysis can be repeated for a freeze-out scenario. In this case the DM particle X is in thermal equilibrium with the SM bath. As the temperature falls below the mass $m_X$, DM particles annihilate into SM particles and the DM abundance is given by the equilibrium abundance of a non-relativistic particle 
\begin{align}
    \rho_{\mathrm{DM,eq}}\propto m_X T^3\left(\frac{m_X}{T}\right)^{3/2} \exp\left(-\frac{m_X}{T}\right)\,.
\end{align}
Eventually the annihilation process gets quenched, as the rate at which DM particles encounter each other is proportional to the exponentially decreasing abundance.
In this case the DM equation of motion including the heat transfer $Q_{\gamma\to\DM}/H$ reads
\begin{align}
    \rho_{\mathrm{DM}}'+3\rho_{\mathrm{DM}}=-\frac{1}{m_X H}\langle\sigma v\rangle_T\left(\rho_{\mathrm{DM}}^2-\rho_{\mathrm{DM,eq}}^{2}\right),
    \label{eq:freeze-out_eom}
\end{align}
with the temperature-averaged cross section $\langle\sigma v\rangle_T$, which we take as constant.

The evolution of energy densities and curvature perturbations is presented in the bottom row of \cref{fig:curv_evo}. The DM component is shown in purple, and the dashed line shows the equilibrium density in the left panel. The model parameters are chosen such that the freeze-out takes place between $N=1$ and $N=2$, after which the DM component keeps scaling as matter.\footnote{Note that the numerical study still considers separate fluids with a fixed equation of state parameter.} More details on the chosen numerical setup are to be found in \cref{sec:app:numeric_evol}.

In \cref{fig:curv_evo_FO:sub2}, as for the freeze-in, we show the evolution of curvature and isocurvature perturbations. 
At the time of the freeze-out, a finite isocurvature perturbation in the DM fluid (black dashed) emerges, which remains constant after the end of the process when the DM scales purely as matter. Compared to the freeze-in, the curvature perturbation of DM (purple) is lowered compared to the radiation bath (black). The isocurvature has the opposite sign.

To analyze the parametric dependence, we perform the evaluation multiple times for different model parameters as in the freeze-in scenario. The resulting isocurvature perturbations are plotted as a function of the initial DR contribution in the right panel of \cref{fig:scan_fit}. We show the results for three masses and four different rates each. While rate and mass minimally change the slope, 
the induced isocurvature is still strictly linear in the initial DR contribution for one set of $(m_X,\Gamma)$. One reason for the small deviations is the different initial equilibrium density and thus a minimally changed initial weighting of constituents.
Remarkably, the analytic and much more simple toy model showed $\Gamma$-dependence for the freeze-out as well, but not for the freeze-in.

The fit results for $m_X=\{2.5,1,0.5 \}$ with $\Gamma=10^5 $ are
\begin{align}
    S_{\gamma\DM}\approx \{0.57,0.55,0.53 \}\,\Omega_{\DR,0}\,S_{\gamma\DR,0}\,.
    \label{eq:slope_FO}
\end{align}

Despite a small dependence of the slope on the model parameters, in general we find the same result as for the freeze-in: The induced DM isocurvature scales linear with the DR abundance during the freeze-out and the amplitude of the initial DR isocurvature perturbation. In contrast to the freeze-in, the slope, while comparable in scale, is positive. The analytic estimate in \cref{app:sudden_freeze} relying on the sudden freeze-out approximation gives slightly larger $\alpha$ but agrees to within 10\%. 

\section{How isocurvature can decipher the origin of DM}
\label{sec:results}

\begin{figure}[t]
\centering
\begin{subfigure}{.5\textwidth}
  \centering
  \includegraphics[width=.9\linewidth]{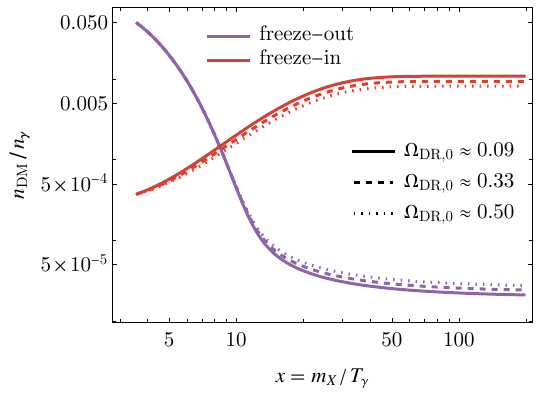}
  \caption{Yield of dark matter component.}
  \label{fig:yield_combi:sub1}
\end{subfigure}%
\begin{subfigure}{.5\textwidth}
  \centering
  \includegraphics[width=.9\linewidth]{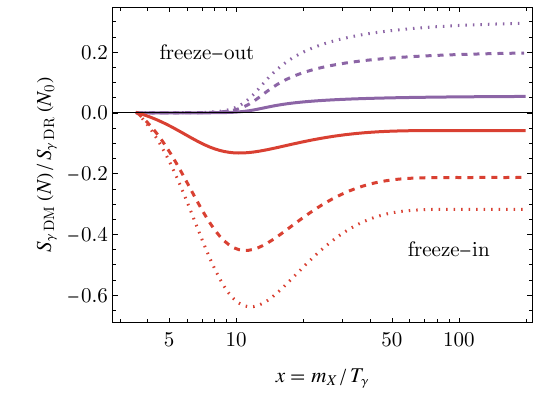}
  \caption{Curvature perturbation evolution.}
  \label{fig:yield_combi:sub2}
\end{subfigure}
\caption{The evolution of the yield $n_{\mathrm{DM}}a^3$, compared to the yield of SM photons $n_{\gamma}a^3$, in freeze-in and freeze-out scenarios is shown for different patches with increasing dark radiation energy density (solid to dotted). Note that for better comparability we choose such values in the numerical setup that the freeze-in happens later on the $x$-axis of the yield plot. This is done by taking a larger $m_X$, which otherwise does not change the setup, since it is degenerate with the choice of $\Gamma$ in the numerical analysis. The right panel demonstrates the evolution of the dark matter isocurvature corresponding to the solid line from the left panel for freeze-in as well as freeze-out. More details on the numerical setup in \cref{sec:app:numeric_evol}.}
\label{fig:yield_combi}
\end{figure}

It was demonstrated above that dark matter freeze-in as well as freeze-out in presence of DR isocurvature from an independent dark sector leads ultimately to nonzero isocurvature in DM. An observation of isocurvature perturbations, on the other hand, could be able to differentiate between the two scenarios. In the case of freeze-in, a higher local DR density increases the Hubble parameter of the patch, which decreases the net energy flow $Q_{1\to2}/H$ and thus leads to later freeze-in and reduced relic abundance. \cref{fig:yield_combi:sub1} shows this reduced relic abundance in the yield (red) for higher background DR densities (increasing from solid to dashed to dotted). Each line can be understood as an independent patch in the separate universe picture.

In freeze-out, the increased Hubble parameter in the patch has the same effect on $Q_{1\to2}/H$, but since the flow is inverse for freeze-out, this implies higher relic abundance. This is shown in \cref{fig:yield_combi:sub1} in purple.
Correspondingly, the induced isocurvature perturbation in \cref{fig:yield_combi:sub2}, sourced by the non-adiabatic energy transfer, acquires a different sign for freeze-in and freeze-out. From an observed isocurvature between photons and DM, the 'direction' of the perturbations could indicate whether freeze-out or freeze-in would be preferred.

\begin{figure}[t]
    \centering
    \includegraphics[width=0.66\textwidth]{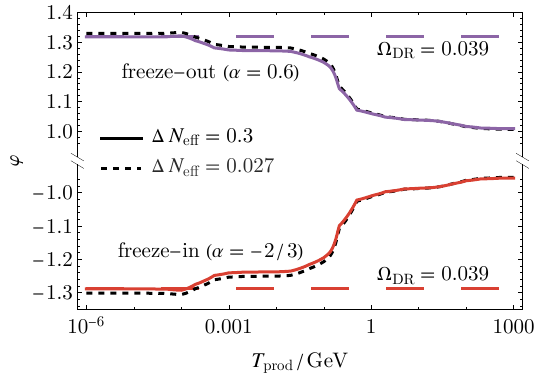}
    \caption{Isocurvature mixing angle as a function of the temperature at DM production, which changes the degrees of freedom in the SM plasma and therefore the relative energy contribution of DR at this time. A freeze-out (purple) leads to positive angles, while freeze-in leads to negative values. We show the results for $\Delta\Neff=0.3$ (solid and long-dashed line) and $\Delta\Neff=0.027$ (short-dashed line). The long-dashed line shows the result if the DR fraction would be constant at the value from recombination.}
    \label{fig:mixing_angle_temp}
\end{figure}

Since the SM neutrinos were in equilibrium with the photons at the time they decouple in all cosmologies we considered, we will assume that they themselves have no isocurvature. We identify DR isocurvature with what we called effective neutrino isocurvature above, 
\begin{align}
    S_{\gamma\nueff}=\frac{\rho^\rec_\DR}{\rho^\rec_{\nueff}}~ S_{\gamma\DR}\,
    \label{eq:S_nueff}
\end{align}
where $\rho^\rec_{\nueff}=\rho^\rec_\nu+\rho^\rec_\DR$. The $\rec$ superscript here indicates that the quantity should be taken at the present day value relevant for CMB and LSS observations. 

We may summarize \cref{eq:slope_FI} and \cref{eq:slope_FO} as $S_{\gamma\DM}=\alpha \Omega^\text{prod}_{\DR} S_{\gamma \DR}$, where $\alpha$ is the $\mathcal{O}(1)$ factor encoding the respective DM production mechanism. Note that for the gravitational effect we described above, the DR density at the time of DM production $\Omega^\text{prod}_\DR$ is of relevance. We may express the mixing angle between neutrino and DM isocurvature as
\begin{align}
    \varphi=\arctan\left(\frac{S_{\gamma\nueff}}{S_{\gamma\DM}}\right)=
    \arctan\left( \frac{1}{\alpha } 
    \frac{
        \rho^\rec_\DR / \rho^\rec_\text{rad}
    }{\Omega^\text{prod}_{\rm DR}}  \frac{\rho^\rec_\text{rad}}{\rho^\rec_{\nueff}}\right)\,,
    \label{eq:mixing_angle_ch4}
\end{align}
where $\rho_\text{rad}=\rho_\gamma+\rho_{\nueff}$\,. The second factor in this parametrization is unity if both the dark radiation and SM ($\gamma,\nu$) sector both exactly redshift like radiation after DM production. This is however improbable as the SM deviates from this scaling when its relativistic degrees of freedom change. One could also imagine more drastic deviations, if for example one of the sectors undergoes an intermediate matter scaling.

Let us now consider a minimal scenario to investigate the impact from bounds on extra relativistic degrees of freedom and thus on DR at recombination. For this, we include the parametrization of $\Delta\Neff$ to later insert the available bounds.
At recombination, the contribution of DR to the total free-streaming radiation $\rho_{\nueff}$ is
\begin{align}
    \frac{\rho_\DR}{\rho_\DR+\rho_\nu}\bigg\rvert_\rec=\frac{\Delta\Neff}{\Delta\Neff+\Neff^{\SM}}
    \approx \frac{\Delta\Neff}{\Neff^{\SM}}+\mathcal{O}\left(\frac{\Delta\Neff}{\Neff^{\SM}}\right)^2
    \,,
    \label{eq:DRtonueff_rec}
\end{align}
which fixes the prefactor from \cref{eq:S_nueff}. This still leaves open the DR contribution during DM production.

Comparing the DR contribution to all radiation in the SM bath yields
\begin{align}
    \frac{\rho_\DR}{\rho_\gamma+\rho_\nu}\bigg\rvert_\rec
    =\frac{\frac{7}{8}\left(\frac{4}{11}\right)^{4/3} \Delta\Neff}{1+\frac{7}{8}\left(\frac{4}{11}\right)^{4/3} \Neff^{\SM}}\,,
\end{align}
which can be computed at a temperature $T$ by adjusting for the (entropy) degrees of freedom in the SM radiation bath at $T$ compared to recombination. Including this, at linear order in $\Delta\Neff$, the DR contribution to the background energy density at $T$ then is 
\begin{align}
    \Omega_\DR(T)&=\frac{\rho_\DR}{\rho_\DR+\rho_\gamma+\rho_\nu}
    \approx 
    \frac{\frac{7}{8}\left(\frac{4}{11}\right)^{4/3}\left(\frac{g_{\star,s}(T_{})}{3.91}\right)^{4/3}\left(\frac{3.38}{g_{\star}(T_{})}\right)}{1+\frac{7}{8}\left(\frac{4}{11}\right)^{4/3}\Neff^{\SM}}\Delta\Neff
    \,,
    \label{eq:OmegaDR_T}
\end{align}
where the prefactor in front of $\Delta\Neff$ in the first order expansion can range from $0.13$ for low temperatures up to $0.35$ above $T\sim200\;\mathrm{GeV}$.

Let us put emphasis on the fact that in \cref{eq:mixing_angle_ch4}, due to the linear behavior of \cref{eq:DRtonueff_rec} and \cref{eq:OmegaDR_T} for small $\Delta\Neff$, in first order the actual amount of DR cancels from the mixing angle. \cref{eq:mixing_angle_ch4} consequently only depends on $\alpha$, $\Neff^\SM$, and the SM degrees of freedom at the time of DM production.

For a realistic estimate let us take into account the upper bound for additional ultra relativistic degrees of freedom \cite{Planck:2018vyg},
\begin{align}
    \Delta\Neff={\frac{8}{7}\left(\frac{11}{4}\right)^{4/3}} \,\frac{\rho_\DR}{\rho_\gamma}<0.3\,,
\end{align}
and assume that DR saturates this limit. At recombination, this fixes \cref{eq:DRtonueff_rec} to $0.09$. In \cref{fig:mixing_angle_temp} we compare the resulting temperature dependence of the mixing angle with one saturating the forecast for CMB-S4 and similar setups of $\Delta\Neff=0.027$ \cite{CMB-S4:2016ple}. Here we set $\alpha= -2/3$ and $\alpha\approx 0.6$ for the freeze-in and freeze-out scenario, respectively. Note that here we are neglecting a possible temperature dependence of $\alpha$ in the freeze-out scenario. As expected the influence of $\Delta\Neff$ on the mixing angle is small, while there is a sizable temperature dependence due to the degrees of freedom in the SM changing, which leads to $\Omega_\DR$ varying. 

\begin{figure}[t]
    \centering
    \includegraphics[width=0.7\textwidth]{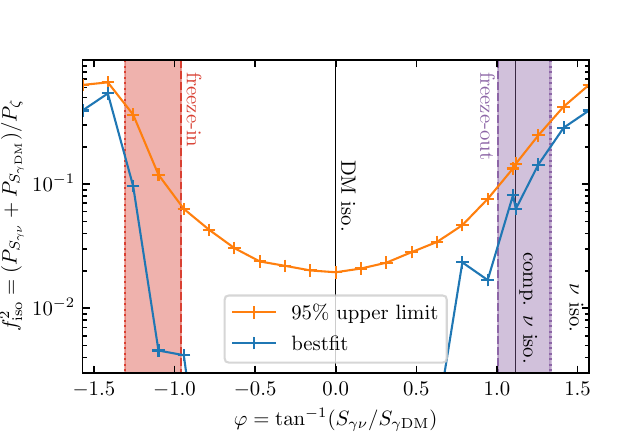}
    \caption{Isocurvature mixing angle, which parameterizes the relative amplitude of effective neutrino and DM isocurvature, that would be expected for freeze-out (purple) and freeze-in (red) scenarios. The band gives the range of possible DR contribution in the low-$\Delta\Neff$ limit, which depends on the temperature at DM production. The orange and blue line show upper limit and best fit found in Ref. \cite{Gerlach:2025hxw} for a common cosmological data set. The angle 0 would be pure DM isocurvature, while $\pm\pi/2$ would be pure neutrino isocurvature. 
    The compensated neutrino isocurvature is the definition tested commonly in the literature (see \cite{Gerlach:2025hxw}).}
    \label{fig:mixing_angle}
\end{figure}

As we demonstrated in~\cite{Gerlach:2025hxw}, the limits on the isocurvature power spectrum from Planck \cite{Planck:2018jri} do exhibit a strong dependence on the mixing angle $\varphi$, with the limits varying by an order of magnitude, as shown in \cref{fig:mixing_angle}. The range of the isocurvature mixing angles expected for freeze-in and freeze-out DM production is indicated by two colored bands. The outer and inner values correspond to the minimal and maximal DR contribution at DM production (cf. \cref{fig:mixing_angle_temp}). This leaves the range $-1.30<\varphi<-0.95$ for freeze-in and $1.03<\varphi<1.34$ for freeze-out. Quite remarkably, the fit has a mild preference for non-vanishing isocurvature of either sign and close to maximal mixing angle, consistent with the range expected for DM production.

We want to note that giving up on the assumption that the DR sector strictly redshifts as radiation would ultimately widen the range of possible mixing angles. For example if the degrees of freedom in the DR sector decrease significantly or if it undergoes an intermediate scaling like matter, values of $\varphi$ closer $\pi/2$ and $-\pi/2$ can be achieved that our fit seems to prefer. 

For the two DM production mechanisms we have discussed here, there is no overlap between the possible mixing angles even when allowing for exotic scalings of the DR sector, since the two values of $\alpha$ and, therefore, the resulting mixing angles have different signs. Given the large amount of possible DM generation scenarios, it is however obvious that in general there is a degeneracy between the value of $\alpha$ of the production mechanism and the temperature at which it takes place. 

Further note that at linear order CMB and large scale structure probes can not distinguish baryon and DM isocurvature \cite{Grin:2011tf,Grin:2011nk}. Since baryogenesis is necessarily a non-equilibrium process \cite{Sakharov:1967dj}, presumably also baryon isocurvature is produced in the cosmologies we discussed here. Similar to the case of DM production, the baryon abundance is determined by competition of various particle physics rates with the Hubble rate and therefore fluctuations in the DR sector would change the yield just as in the DM case. The total value of $\alpha$ encoding the correlation between neutrino/DR isocurvature and the total matter curvature then depends both on the DM production and baryogenesis mechanism.  

{
As we have seen, the combination of isocurvature modes, especially the sign of the mixing angle, provide the main signature for such cosmological histories as discussed here. In \cite{Gerlach:2025hxw} we demonstrated that the sign of the mixing angle can significantly change the power spectrum of CMB anisotropies and matter perturbations. While the fit produced there has no preference for one sign of $\varphi$, this could be one way to engage the ambiguity in future. 
}

{
Isocurvature sources different from primordial fields, such as first-order phase transitions \cite{Buckley:2025zgh}, formation of primordial black holes \cite{Gerlach:2025vco,Franciolini:2026fdv}, or topological effects may exhibit specific spectral shapes, non-Gaussianities \cite{Kawasaki:2008sn}, or strong scale-dependence \cite{Chluba:2013dna,Ramberg:2022irf}, while the primordial origin from light fields implies nearly Gaussian and scale-independent statistics for the perturbations \cite{Gordon:2000hv,
Lyth:2001nq,
Wands:2007bd}.
The correlation to the curvature perturbation depends on the UV-history. In the simplest setup no correlation between the isocurvature and the curvature mode would be expected. The case of no correlation is also preferred by the fit performed in \cite{Gerlach:2025hxw}.
}

\section{Discussion}

In this work we have derived a closed expression in \cref{eq:evolution_zeta_general} for the superhorizon evolution of cosmic perturbations within the separate universe picture.
We applied this formula to the evolution of isocurvature perturbations during DM production in  presence of a dark radiation component. 
Since isocurvature can not be created from adiabatic initial conditions, the DR must necessarily get reheated by a different field than the SM bath, such that it features isocurvature to begin with.
We found that generically DM isocurvature is induced even if the dark radiation is only coupled gravitationally.
The reason is that local variations in the DR background will modify the Hubble rate  and therefore the efficiency of the DM production process. 
Depending on the direction of energy transfer—whether the DM is freezing-out or freezing-in—the yield may increase or decrease in a patch with higher dark radiation background density, and vice versa. The sign of this energy flow determines the sign of the isocurvature perturbation imprinted on DM.

In a numerical study as well as in analytic estimates we found that for both freeze-in and freeze-out scenarios, the DM isocurvature scales linearly with the product of DR energy contribution to the background and the initial isocurvature perturbation of DR, with a prefactor $\alpha$ of $\mathcal{O}(1)$ that depends on the production mechanism. For direct freeze-in we found no dependence on microscopic DM parameters, for freeze-out it has a very mild temperature dependence.
We expect the result to only break down for scales which enter the horizon while the process is not completed.

Future work can be conducted in two directions.
First, the framework can be extended to more realistic cosmological histories and particle models. Determining how the slope $\alpha$ actually varies for different mechanisms of 
DM production like, e.g., UV-freeze-in could help to lift the degeneracy with the temperature at which the process takes place.

For the minimal cosmic history, 
a promising scenario would be the observation of a finite $\Delta \Neff$. Such a result would motivate a more detailed search for mixed isocurvature modes, since an imprint on DM isocurvature would be likely. 
If cosmological data shows preference for a nonzero isocurvature amplitude and a value for $\varphi$, our framework will be able to discriminate between freeze-in and freeze-out scenarios. Either one of the $\alpha$-values for freeze-in or freeze-out of the standard model sector is preferred, or DM is created non-minimally.
If, on the other hand, the preference remains close to pure neutrino isocurvature—in line with our previous result \cite{Gerlach:2025hxw}, the then decisive question would be to explain why the imprint into DM isocurvature is so small, since the freeze-in/-out bands do not coincide with the mixing angle being close to $\pm\pi/2$ (cf. \cref{fig:mixing_angle}). The measurements could as well shed light on other non-equilibrium processes such as baryogenesis, which similarly depend on the local Hubble rate.

So far, we have only considered scenarios where the DR behaves strictly as radiation.
A more detailed exploration of scenarios with nontrivial DR evolution, such as the recently proposed \textit{phase-in} mechanism \cite{Benso:2025vgm}, could find more efficient ways to induce DM isocurvature. 
The only condition is, that there is a separate sector with isocurvature, which is relevant or even dominant during the DM production. Another alternative is that DM freezes out of DR rather than out of the SM bath.

In summary, the connection between DR isocurvature and the origin of DM opens a new window to probe DM microphysics via cosmological data, and may hold a key clue to solving the dark matter puzzle.

\acknowledgments{The authors thank Kim V. Berghaus and Davide Racco for insightful discussions on dark matter isocurvature. PS and CG acknowledge support by the Cluster of Excellence “Precision Physics, Fundamental Interactions, and Structure of Matter” (PRISMA+ EXC 2118/1) funded by the Deutsche Forschungsgemeinschaft (DFG, German Research Foundation) within the German Excellence Strategy (Project No. 390831469).}


\newpage
\appendix

\section{Cosmic Perturbation Theory}
\label{app:full_metric}
The general form for metric perturbations at first order in a spatially flat Friedmann–Lemaître– Robertson–Walker (FLRW) metric background is \cite{Mukhanov:1990me}
\begin{align}
    ds^2=-(1+2\phi)dt^2+2a B_{,i}dt dx^i+a^2\left[(1-2\psi)\delta_{ij}+E_{,ij}\right]dx^idx^j~,
\end{align}
where $t,x$ denote the coordinates and $a$ the scale factor. 
$B_{,i}$ can be decomposed as a scalar $B$ and a divergenceless vector $B_i$,
$B_{,i}=\partial_i B+B_i$. Similarly, one can decompose $E_{,ij}$
into a scalar part $E$, a divergenceless vector $E_i$, and the tensor $h_{ij}$, which is divergenceless and traceless, 
\begin{align}
    E_{,ij}=2 \partial_i \partial_j E + ( \partial_i E_j + \partial_j E_i) + h_{ij}\,.
\end{align}
Since we are not interested in vector and tensor perturbations here, we neglect those.
This leaves us with four scalar perturbations, $\phi$, $B$, $\psi$, and $E$. However, they still depend on the slicing and threading of spacetime.

To circumnavigate the gauge problem, one can either choose gauge invariant combinations of the perturbations as the Bardeen variables \cite{Bardeen:1980kt}—or fix the gauge and thereby decide for a specific choice of threading and slicing. The consequence of fixing the gauge is that automatically matter perturbations are generated. One of the easiest choices is the Newton gauge, where two scalar perturbations (here $E$ and $B$) are set to zero. We are left with $\phi$ and $\psi$. The perturbed metric with two scalar perturbations is the one we consider in \cref{eq:Newtonian_metric}.

\subsection{General coordinate transformation}

Consider a coordinate transformation
\begin{align}
    x^\mu\mapsto \tilde{x}^\mu= x^\mu+\xi^\mu(t, \bm{x})\,,
\end{align}
where the change can be decomposed\footnote{We already ignore a possible divergenceless vector perturbation $\beta^i$ in the spatial part here, since we are only interested in the scalar perturbations.} into $\xi^\mu=(\alpha,\partial^i \beta)$. Under this transformation, the scalar perturbations above transform as
\begin{gather}
    \phi \mapsto \tilde{\phi}= \phi-\dot{\alpha}\,, \\
    B\mapsto \tilde{B}= B+\frac{\alpha}{a}-a\dot{\beta}\,,\\
    \psi\mapsto \tilde{\psi}= \psi+\alpha H\,,\\
    E\mapsto \tilde{E}= E-\beta \,.
\end{gather}
As a consequence, if we consider a time shift $\delta t$, but no shift in space, thus $\xi^\mu=(\delta t,0)$, the scalar perturbation in the diagonal spatial part transforms as
\begin{align}
    \psi\to \psi + H\delta t\,.
\end{align}

\subsection{Transforming energy density perturbations}
Next we consider a general scalar field. Without loss of generality we take some energy density $\rho(t,\bm{x})$. A scalar field under the general coordinate transformation transforms as $\tilde{\rho}(\tilde{t},\tilde{\bm{x}})=\rho(t,\bm{x})$. 
This implies for the infinitesimal transformation given by $\xi^\mu$ at first order
\begin{equation}
    \tilde{\rho}(\tilde{t},\tilde{\bm{x}})=\rho(t,\bm{x})=\rho(\tilde{t}-\alpha,\tilde{\bm{x}}-\nabla \beta) \approx \rho(\tilde{t},\tilde{\bm{x}})-\alpha\dot{\rho} (\tilde{t},\tilde{\bm{x}})-\partial_i\rho(\tilde{t},\tilde{\bm{x}}) \partial^i \beta\,.
\end{equation}
In general, we assume the background to be only dependent on the time, $\bar{\rho}(t)$. We split the transformed field into background plus a perturbation and find (now dropping the tilde on the new coordinates)
\begin{align}
    \tilde{\rho}(t,\bm{x})=\bar\rho(t)+\delta\tilde\rho(t,\bm{x}) \overset{!}{=} \bar\rho(t) +\delta\rho(t,\bm{x})-\alpha\dot{\bar\rho} (t)~,
\end{align}
where the spatial derivative is left out since its application to the perturbation $\delta\rho$ is already a second order perturbation. Note that we derived the transformation of an energy density perturbation,
\begin{equation}
    \delta\tilde\rho=\delta \rho-\dot{\bar\rho}\,\delta t\,.
\end{equation}

\section{Changing the time variable}
\label{app:time_coordinate_change}

The choice of the time variable in principle is free. Throughout this article, we have used cosmic time $t$ as well as e-folds $N$. 
If $t$ now is replaced by a time reparameterization along the background trajectory, $\tau=\tau(t,\bm{\rho}_0)$, 
additional contributions in \cref{eq:evolution_zeta_general} cancel, as we will see now.

Let us first note that the total differential of $\tau$ reads
\begin{align}
    \upd \tau = \frac{\partial\tau}{\partial t}\upd t+\sum_\gamma \frac{\partial\tau}{\partial\rho_{0,\gamma}}\upd\rho_{0,\gamma}\,.
\end{align}
We now want to examine a variation of $\rho_{0,\beta}$ as in \cref{eq:evolution_zeta_general}. Let us see what happens to the cosmic time when we do this while fixing $\tau$,
\begin{align}
    \frac{\partial t}{\partial\rho_{0,\beta}}\bigg\rvert_\tau=-\frac{\frac{\partial\tau}{\partial\rho_{0,\beta}}}{\frac{\partial\tau}{\partial t}}
\end{align}
since by fixing $\tau$ we imply $\upd\tau=0$.
Now we apply this variation to our background fields, $\rho_\alpha(t,\bm{\rho}_0)$ and $N(t,\bm{\rho}_0)$. This yields
\begin{align}
    \frac{\partial \rho_\alpha}{\partial\rho_{0,\beta}}\bigg\rvert_t&=
    \frac{\partial \rho_\alpha}{\partial\rho_{0,\beta}}\bigg\rvert_\tau
    +
    \frac{\partial\rho_\alpha}{\partial t}\bigg\rvert_{\rho_0}\,
     \frac{\partial t}{\partial\rho_{0,\beta}}\bigg\rvert_\tau
     \\
     &=\frac{\partial \rho_\alpha}{\partial\rho_{0,\beta}}\bigg\rvert_\tau
    -
    \dot{\rho}_\alpha
    \left(
    \frac{\partial\tau/\partial\rho_{0,\beta}}
    {\partial\tau/\partial t}
    \right)
    \label{eq:app_time_var_proof_rhs1}
\end{align}
and
\begin{align}
    \frac{\partial N}{\partial\rho_{0,\beta}}\bigg\rvert_t&=
    \frac{\partial N}{\partial\rho_{0,\beta}}\bigg\rvert_\tau
    +
    \frac{\partial N}{\partial t}\bigg\rvert_{\rho_0}\,
     \frac{\partial t}{\partial\rho_{0,\beta}}\bigg\rvert_\tau
     \\
     &=\frac{\partial N}{\partial\rho_{0,\beta}}\bigg\rvert_\tau
    -
    H
    \left(
    \frac{\partial\tau/\partial\rho_{0,\beta}}
    {\partial\tau/\partial t}
    \right)\,,
    \label{eq:app_time_var_proof_rhs2}
\end{align}
where in the last step we replaced $\dot{N}=H$.

With this result we can consult our evolution equation (\cref{eq:evolution_zeta_general}) again, which contains
\begin{align}
    \left[ \frac{H(t)}{\dot\rho_{\alpha}(t)}\frac{\partial \rho_{\alpha}(t)}{\partial\rho_{0,\beta}} - 
    \frac{\partial N(t)}{\partial \rho_{0,\beta}}\right]\,.
    \label{eq:app_evolution_bracket}
\end{align}
Here, the prefactors of the two variations w.r.t. $\rho_{0,\beta}$ make sure that the last terms in \cref{eq:app_time_var_proof_rhs1} and \cref{eq:app_time_var_proof_rhs2} cancel each other, leaving the evolution equation unaltered except for the arguments. This together with the fact that 
\begin{equation}
    \frac{dN/dt}{d\rho_\alpha/dt}=\frac{dN/d\tau}{d\rho_\alpha/d\tau}\,,
\end{equation}
proves our claim that \cref{eq:evolution_zeta_general} can be applied for all such time arguments $\tau$, when interpreting $H=dN/d\tau$ as a generalized Hubble rate.

The e-fold parameterization is a special choice of such a generalized time, as the second term in \cref{eq:app_evolution_bracket}, which varies the initial density while fixing the time parameter $N$, vanishes naturally. Further the generalized Hubble rate with respect to e-folds is $dN/dN=1$\,. As a consequence, the dimensionless choice of $N$ as time parameter yields the simpler form of the evolution equation for $\zeta(N)$ in \cref{eq: zeta(N)}.

\section{Superhorizon evolution of isocurvature in e-folds and cosmic time in comparison}
\label{sec:isocurvature_evolution}
We showed in \cref{eq: evolution delta rho} that the evolution equation for the density fluctuation may be derived from the background evolution. 
We then went on to argue that the time dependence of the perturbations may be found directly by considering solutions of the background with variations of the initial conditions and, therefore, did not further investigate the possibility of deriving and solving evolution equations for the perturbations. 
In \cref{sec:decaying_curvaton} we showed that the results for the time evolution for the perturbations using variation of the initial conditions and solving their equations of motion (from Ref. \cite{Malik:2002jb}) coincide in the case of the decaying curvaton.
We here derive the evolution equations for the perturbations in the case of multiple interacting fluids starting from \cref{eq: evolution delta rho} and without going through cosmic perturbation theory explicitly as was done in Ref. \cite{Malik:2002jb}.

\subsection{The evolution with respect to e-folds}
Since shifts in e-folds correspond to just a rescaling of the scale factor, the background equations are also autonomous when working with e-folds. We may, therefore, start from \cref{eq: evolution delta rho} but using e-folds $N$ rather than physical time $t$\,, which we restate here for convenience,
\begin{align}
    \delta\rho'_\alpha=\sum_\beta \delta\rho_\beta \frac{\partial \tilde{F}_\alpha(\bm{\rho}(N))}{\partial \rho_\beta}\,,
\end{align}
where as before primes denote derivatives with respect to e-folds. The tilde indicates that this relates to the background evolution in e-folds $\bm{\rho}'=\tilde{\bm F}(\bm \rho)$. The benefit of using e-folds here is the same as in \cref{sec: curvature and isocurvature perturbations}: We may use a slicing such that the metric perturbation $\psi=0$ vanishes at $N_0$\,. It will then vanish at all $N=N_0+\Delta N$, such that the gauge invariant perturbation $\zeta_\alpha$ reduces to $\zeta_\alpha=-\delta\rho_\alpha/\rho_\alpha'$\,.

We can then compute
\begin{align}
    \zeta_\alpha'=-\ddN{}\left( \frac{\delta\rho_\alpha}{\rho_\alpha'} \right) 
    &=\frac{1}{\rho_\alpha'} \sum_\beta -\delta\rho_\beta \frac{\partial \tilde{F}_\alpha(\bm{\rho}(N))}{\partial \rho_\beta}+\frac{\delta\rho_\alpha}{\left(\rho_\alpha'\right)^2}\ddN{\tilde{F}_\alpha(\bm{\rho}(N))}
    \label{eq:app:zeta_eom}
    \\
    &= 
    -\frac{1}{\rho_\alpha'}
    \sum_\beta \frac{\partial \tilde{F}_\alpha(\bm{\rho}(N))}{\partial\rho_\beta}\left(\delta\rho_\beta- \frac{\delta\rho_\alpha}{\rho_\alpha'}\,\rho_\beta'\right)\,,
    \label{eq:ddN_dens_pert_fraction}
\end{align}
where in the second step the derivative of $\tilde{F}_\alpha(\rho_i)$ is expanded.

\subsection{Evolution for interacting fluids with respect to time}
Let us now consider the special case of fluids with some heat transfer $Q_\alpha$ between them as in \cite{Malik:2002jb}.
The equation of motion for fluid $\alpha$ can be stated as
\begin{align}
    \rho'_\alpha=\tilde{F}_\alpha(\bm{\rho})=-3(\rho_\alpha+P_\alpha(\bm{\rho}))+\Tilde{Q}_\alpha(\bm{\rho})\,,
\end{align}
where pressure $P_\alpha$ denotes the pressure of fluid $\alpha$ and $\Tilde{Q}_\alpha$  the energy transfer (w.r.t. $N$). Both are functions only depending on the various energy densities of all present fluids (we suppress the arguments below). Therefore, we can replace the term
\begin{align}
\label{eq:app:partial_F}
    \frac{\partial \tilde{F}_\alpha}{\partial\rho_\beta}=-3\delta_{\alpha \beta}-3\frac{\partial P_\alpha}{\partial\rho_\beta}+\frac{\partial \Tilde{Q}_\alpha}{\partial\rho_\beta}
\end{align}
and find 
\begin{align}\label{eq:app:zeta_eom_sum}
    \zeta'_\alpha
    &=-\frac{1}{\rho'_\alpha}\left[
        -3\sum_\beta \frac{\partial P_\alpha}{\partial\rho_\beta} \left(  \delta\rho_\beta - \frac{\delta\rho_\alpha}{\rho'_\alpha}\,\rho'_\beta\right)
        +\sum_\beta \frac{\partial \Tilde{Q}_\alpha}{\partial\rho_\beta} \left(  \delta\rho_\beta - \frac{\delta\rho_\alpha}{\rho'_\alpha}\,\rho'_\beta\right)
    \right]\,.
\end{align}
Note that the delta in \cref{eq:app:partial_F} cancels when contracted with the parenthesis on the right of \cref{eq:ddN_dens_pert_fraction}. 

Let us now go from e-folds to regular time in order to be able to compare our result with \cite{Malik:2002jb}. To do so we replace $d/d N=H^{-1}\,d/d t$ as well as $\Tilde{Q}_\alpha=Q_\alpha/H$\,, where $Q_\alpha$ is the heat transfer to fluid $\alpha$ per time.

The result may be written as
\begin{align}
    \label{eq:app:zeta_eom_final}
    \dot\zeta_\alpha&=
    \frac{H}{\dot{\rho}_\alpha}\left[
        3H\delta P_{\mathrm{intr.,}\alpha}
        -\delta Q_{\mathrm{intr.,}\alpha}-\delta Q_{\mathrm{rel.,}\alpha}
    \right]\,.
\end{align}
The first term results from the variation of $P_\alpha$ in \cref{eq:app:zeta_eom_sum},
\begin{align}
    \delta P_{\mathrm{intr.,}\alpha}&=\sum_\beta \frac{\partial P_\alpha}{\partial\rho_\beta} \left(  \delta\rho_\beta - \frac{\delta\rho_\alpha}{\dot\rho_\alpha}\,\dot\rho_\beta\right)\\
    &=\sum_\beta \frac{\partial P_\alpha}{\partial\rho_\beta}  \delta\rho_\beta - c_\alpha^2\,\delta\rho_\alpha\,,
\end{align}
where $c_\alpha$ denotes the speed of sound in fluid $\alpha$\,, $c_\alpha^2=\dot P_\alpha/\dot \rho_\alpha$\,. 

For the second sum, we first replace $\Tilde{Q}_\alpha=Q_\alpha/H$ and realize that
\begin{align}
    \frac{\partial}{\partial\rho_\beta}\left(\frac{Q_\alpha}{H}\right)=\frac{1}{H}\left(\frac{\partial Q_\alpha}{\partial\rho_\beta}-\frac{Q_\alpha}{2\rho}\right)\,,
\end{align}
since $\partial H/\partial\rho_\beta=\partial H/\partial\rho=H/(2\rho)$, where we used that the total density $\rho$ is the sum of all $\rho_\beta$ and $H\propto\sqrt{\rho}$ in the Friedmann equation. 

Plugging this result back into \cref{eq:app:zeta_eom_sum}, we get one term due to the variation of $Q_\alpha$ that may be written as
\begin{align}
    \delta Q_{\mathrm{intr.,}\alpha}
    &=\sum_\beta \frac{\partial Q_\alpha}{\partial\rho_\beta}  \delta\rho_\beta - \frac{\dot Q_\alpha}{\dot\rho_\alpha}\,\delta\rho_\alpha\,
\end{align}
in analogy to $\delta P_{\mathrm{intr.,}\alpha}$. 
The remaining term can be written as
\begin{align}
    \delta Q_{\mathrm{rel.,}\alpha}&=-\frac{Q_\alpha}{2\rho}\sum_\beta  \left(  \delta\rho_\beta - \frac{\delta\rho_\alpha}{\rho'_\alpha}\,\rho'_\beta\right)\\
    &=\frac{Q_\alpha}{2\rho}\left(\frac{\dot\rho}{\dot\rho_\alpha}\delta\rho_\alpha-\delta\rho\right)\,.
\end{align}
With \cref{eq:app:zeta_eom_final} we therefore find the same result as Ref.~\cite{Malik:2002jb} for the evolution of the curvature perturbations.

\section{Setup for numerical evolution}
\label{sec:app:numeric_evol}

In this section we comment on the parameter range and difficulties of the numerical study in \cref{subsec:FI_numerical} and \cref{subsec:FO_numerical}. First of all, we do not give units above. Everything with mass dimension can be assumed to have the same energy unit, say GeV. To simplify the numerics, we normalize the initial total energy density to $\mathcal{O}(1\,\mathrm{GeV}^4)$ and set the reduced Planck mass to $\mPl=1\,\mathrm{GeV}$. This gives initial temperature and Hubble of $\mathcal{O}(1\,\mathrm{GeV})$ as well, making it easier to compare the toy model parameters and setting the time-axis to a few e-folds. 

In \cref{fig:curv_evo} we use $m_B=0.25$, $\Gamma=10^{-5}$, and $\rho_\gamma=1$ for the freeze-in and $m_X=1.5$ with $\Gamma=5\times10^6$ for the freeze-out. Note that we normalize the freeze-in interaction to the initial $\Gamma$. The other prefactors in \cref{eq:Q_freeze-in} therefore do not matter for the toy model. The result for freeze-in does not depend on $m_X$, as it just enters the energy transfer as a constant factor. For better visibility in the yield plot, however, we choose the value of $m_X$ such that the freeze-in happens at a later $x$ than usually assumed ($m_X=1.5$ for \cref{fig:curv_evo}). The initial densities can be read off. The same parameters with different DR contributions as indicated are used in \cref{fig:yield_combi}. For all setups in the context of DM we use the initial curvature perturbations $\zeta_{\gamma}(N_0)=\zeta_{\DM}(N_0)=-\zeta_{\DR}(N_0)=0.1$, which, due to the linear order in which we work, can always be renormalized later to realistic values.

It is possible that parts of the curvature perturbations and, as a consequence, the corresponding isocurvature, can acquire a divergence, when one $\rho_i'$ has a root. Nonetheless, this is not a problem, since $\zeta$ is not a direct observable and the total curvature perturbation will stay smooth. In that case one can always find a combination of curvature perturbations to avoid these poles.
The initial DM abundance is set in the freeze-in to a small non-zero value. For the background plot in \cref{fig:curv_evo} we use an initial DM abundance of $10^{-5}$ to avoid these divergences in the given example.
To avoid influence from initial conditions in the linear fit, we decrease the initial DM abundance for the freeze-in fits, where we only show the final values of the evolution. The initial DM abundance in the freeze-out case is matched to the equilibrium abundance. 

\begin{figure}
    \centering
    \includegraphics[width=0.66\textwidth]{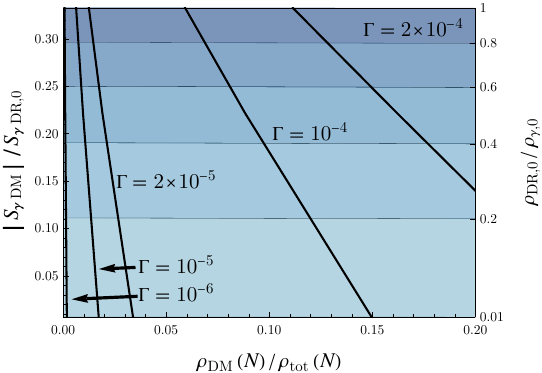}
    \caption{The parameter space for isocurvature in freeze-in dark matter. We plot the final isocurvature as a function of the final dark matter energy density at the end of the evaluation. For the black lines, the decay rate parameter is fixed and the initial dark radiation contribution is varied. For the horizontal lines distinguishing the background color, the rate is varied and the DR contribution fixed. The horizontal lines show that the isocurvature is strictly linear in the initial DR density.}
    \label{fig:scan_FI_parameters}
\end{figure}

For the linear fit to the freeze-in given in \cref{fig:scan_sub1} we use the same progenitor mass as in \cref{fig:curv_evo} and vary $\rho_\DR(N_0)$ from $0.1$ to $1$, which indicates the x-axis. The rate is varied over five discrete values from $\Gamma=10^{-4}$ to $10^{-7}$, but the points cannot be distinguished for different rates. How the resulting DM density changes with the rate is depicted in \cref{fig:scan_FI_parameters}. The setup is the same as in the fit, but the x-axis is given by the final DM density. The colors show the initial DR density. The lines result from fixing a rate each and varying the initial DR density of the setup.

\begin{figure}
    \centering
    \includegraphics[width=0.66\textwidth]{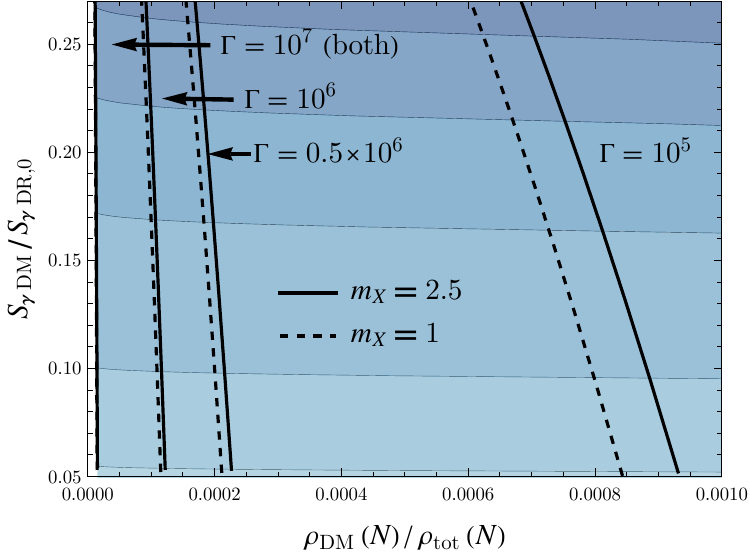}
    \caption{The parameter space for isocurvature in freeze-out dark matter. Shown is the final isocurvature as a function of the final dark matter energy density at the end of the evaluation. The black lines correspond to fixed rates, solid and dashed for different DM masses. The (more or less) horizontal lines are for fixed initial DR contribution and varying the rate, which gives a different final DM density, but roughly the same isocurvature for the range of rates covered.}
    \label{fig:scan_FO_parameters}
\end{figure}

The freeze-out fit in \cref{fig:scan_sub2} is done for three different masses. The fit is only applied to the rate $\Gamma=10^5$ as indicated (solid, dashed and dotted line for $m_X=1,\,2.5,\,0.5$, respectively). We show the points for three masses and four rates. The slope increases slowly if the mass or the rate is increased. As for the freeze-in, we show the parameter space with the final DR density on the x-axis in \cref{fig:scan_FO_parameters}. Here, two masses with four rates each are shown. The horizontal contour lines refer to fixed DR contributions of $\{0.2,0.4,0.6,0.8 \}$.
For all evolutions where the final value is read out this happens at $N=4$ e-folds.

\section{Sudden freeze-in/-out approximation}
\label{app:sudden_freeze}
We here adopt the analysis of \cite{Lyth:2003ip} from the curvaton to the DR isocurvature case to find analytic estimates for the induced DM isocurvature.\\

\textbf{Freeze-in.}
The sudden freeze-in approximation relies on the observation that DM production is most efficient when the SM bath temperature matches the mass of the mediating particle, $T=m_B$\,. The relic abundance is therefore given by
\begin{align}
    \rho_\DM\approx \left.\frac{Q_{\gamma\to\DM}}{H}\right|_{T=m_B}\left(\frac{a}{a_{T=m_B}}\right)^{-3}\,,
\end{align}
while the SM and DR density are simply determined by redshifting
\begin{align}
    \rho_\gamma=\rho_{\gamma,0}\left(\frac{a}{a_{0}}\right)^{-4}\,,\quad \rho_\DR=\rho_{\DR,0}\left(\frac{a}{a_{0}}\right)^{-4}\,.
\end{align}
Note that the scaling of the SM sector $\gamma$ as $a^{-4}$ is only approximate, as at the time of freeze-in the mediating particle $B$ becomes non-relativistic leading to a change of the effective degrees of freedom. Further some of its energy is transferred to the DM sector. Working with e-folds as our time-coordinate $a$ and $a_0$ are independent of the initial conditions and the only non-trivial dependence is encoded by $a_{T=m_B}$ and $H_{T=m_B}$, while $Q_{\gamma\to\DM}(T=m_B)$ is a constant. From the scaling of the SM density we find
\begin{equation}
    a_{T=m_B}=a_0\left(\frac{\rho_{\gamma,0}}{\rho_{\gamma, T=m_B}}\right)^{1/4}\,.
\end{equation}
For the Hubble rate one finds
\begin{equation}
    H_{T=m_B}\propto\sqrt{\rho_{\gamma,T=m_B}+\rho_{\DR,T=m_B}}\propto\sqrt{1+\rho_{\DR,T=m_B}/\rho_{\gamma,T=m_B}}\propto\sqrt{1+\rho_{\DR,0}/\rho_{\gamma,0}}\,.
\end{equation}
Using \cref{eq:evolution_zeta_general} we may now calculate the induced DM isocurvature
\begin{align}
    S_{\gamma\DM}&=\sum_{\mu,\nu=\{\gamma,\DR\}}\left[\frac{1}{4}\frac{\partial\log\rho_{\gamma}}{\partial\log\rho_{\mu,0}}-\frac{1}{3}\frac{\partial\log\rho_{\DM}}{\partial\log\rho_{\mu,0}}\right]4\Omega_{\nu}S^0_{\mu\nu}\\
    &=\left(1-\frac{4}{3}\frac{\partial\log\rho_{\DM}}{\partial\log\rho_{\gamma,0}}\right)\Omega_\DR S_{\gamma \DR}+\frac{4}{3}\frac{\partial\log\rho_{\DM}}{\partial\log\rho_{\DR,0}}\Omega_\gamma S_{\gamma \DR}\label{eq: dark matter isocurvature sudden approximation}
\end{align}
The term arising due to the dependence of $a_{T=m_B}$ on $\rho_{\gamma,0}$ cancels the 1 in the first bracket. The dependence on $H_{T=m_B}$ finally leads to
\begin{align}
    S_{\gamma\DM}&=-\frac{2}{3}\Omega_\DR S_{\gamma \DR}\,.
\end{align}

\textbf{Freeze-out.} In this case the DM density is approximately given by the equilibrium abundance at the temperature $T_\text{out}$, when the annihilation of DM particles becomes inefficient and takes on matter scaling afterwards
\begin{align}
    \rho_\DM\propto m_X \left(m_X T_\text{out}\right)^{3/2}\exp\left(-\frac{m_X}{T_\text{out}}\right)\left(\frac{T}{T_\text{out}}\right)^{3}
\end{align}
Again we will approximate the energy density in the SM and DR to scale as $a^{-4}$ as well as $T\propto \rho_\gamma^{1/4}$\,. We may then use \cref{eq: dark matter isocurvature sudden approximation} to find the isocurvature. The dependence of the dark matter density on the final temperature cancels the 1 and we are left only with the non-trivial dependence due to $T_\text{out}$
\begin{align}
    S_{\gamma\DM}&=\frac{4}{3}\frac{\partial\log\rho_{\DM}}{\partial\log T_\text{out}}\left[-\frac{\partial\log T_\text{out}}{\partial\log\rho_{\gamma,0}}\Omega_\DR+\frac{\partial\log T_\text{out}}{\partial\log\rho_{\DR,0}}\Omega_\gamma\right]S_{\gamma \DR}\\
    &=\frac{4}{3}\left(-\frac{3}{2}+\frac{m_X}{T_\text{out}}\right)\left[-\frac{\partial\log T_\text{out}}{\partial\log\rho_{\gamma,0}}\Omega_\DR+\frac{\partial\log T_\text{out}}{\partial\log\rho_{\DR,0}}\Omega_\gamma\right]S_{\gamma \DR}\,.
\end{align}
The point of freeze-out is reached when the interaction rate per DM particle $\Gamma=\langle\sigma v\rangle_T n_\DM$ drops below the Hubble rate. As in the main text, we will assume that the averaged cross-section does not depend on the temperature. Up to $T_\text{out}$, the dark matter abundance may be approximated by the equilibrium such that $\Gamma_\text{out}\propto T_\text{out}^{3/2}\exp(-m_X/T_\text{out})$\,. Similar to the freeze-in case we may rewrite the Hubble rate at freeze-out as
\begin{align}
    H_\text{out}\propto \sqrt{\rho_{\gamma,\text{out}}+\rho_{\DR,\text{out}}}\propto\frac{T_\text{out}^2}{\sqrt{\rho_{\gamma,0}}}\sqrt{\rho_{\gamma,0}+\rho_{\DR,0}}\,.
\end{align}
From here we may find the dependence of $T_\text{out}$ on the initial SM and DM density by ensuring that $\Gamma_\text{out}/H_\text{out}$ is a constant,
\begin{align}
    0&=d\log\left(\frac{\Gamma_\text{out}}{H_\text{out}}\right)\\
    &=\left(\frac{3}{2}-\frac{m_X}{T_\text{out}}-2\right)d\log T_\text{out}+\left(\frac{1}{2}-\frac{1}{2}\Omega_\gamma\right)d\log\rho_{\gamma,0}-\frac{1}{2}\Omega_\DR d\log\rho_{\DR,0}\\
    &=\left(\frac{m_X}{T_\text{out}}-\frac{1}{2}\right)d\log T_\text{out}+\frac{1}{2}\Omega_\DR d\log\rho_{\gamma,0}-\frac{1}{2}\Omega_\DR d\log\rho_{\DR,0}\,.
\end{align}
We then have in total the induced isocurvature perturbation
\begin{align}
    S_{\gamma\DM}
    &=\frac{2}{3}\frac{m_X/T_\text{out}-3/2}{m_X/T_\text{out}-1/2}\Omega_\DR S_{\gamma \DR}\,.
\end{align}


\bibliographystyle{JHEP}
\bibliography{references}
\end{document}